\documentclass[preliminary,copyright,creativecommons,final]{eptcs}
\providecommand{\event}{GandALF 2022}

\usepackage[utf8]{inputenc}
\usepackage{amssymb}
\usepackage{amsmath}
\usepackage[shortlabels]{enumitem}
\usepackage{xspace}
\usepackage{scalerel}
\usepackage{etoolbox}
\usepackage{stmaryrd}
\usepackage{ifdraft}
\usepackage{xcolor}
\usepackage{wrapfig}

\newtheorem{theorem}{Theorem}
\newtheorem{lemma}{Lemma}
\newtheorem{corollary}{Corollary}

\newtheorem{proposition}{Proposition}
\newtheorem{definition}{Definition}
\newtheorem{example}{Example}
\newtheorem{remark}{Remark}

\usepackage{tikz}
\usetikzlibrary{calc}
\usetikzlibrary{positioning}
\usetikzlibrary{arrows,automata,shapes}

\RequirePackage{centernot}
\RequirePackage[Symbol]{upgreek}

\RequirePackage{etoolbox}
\RequirePackage{mathtools}

\let\originalleft\left
\let\originalright\right
\renewcommand{\left}{\mathopen{}\mathclose\bgroup\originalleft}
\renewcommand{\right}{\aftergroup\egroup\originalright}

\newcommand{\Nat}{\mathbb{N}}

\newcommand{\Procs}{\ensuremath{\mathcal{P}}}
\newcommand{\procA}{\ensuremath{P}}
\newcommand{\procB}{\ensuremath{Q}}
\newcommand{\procC}{\ensuremath{R}}
\newcommand{\procD}{\ensuremath{S}}

\newcommand{\RcvEvs}{\ensuremath{R}}
\newcommand{\SndEvs}{\ensuremath{S}}
\newcommand{\eventnodes}{\ensuremath{N}}
\newcommand{\edges}{\ensuremath{E}}
\newcommand{\val}{\ensuremath{m}}
\newcommand{\vertexA}{\ensuremath{v}}

\newcommand{\BMSCs}{\ensuremath{\mathcal{M}}}

\newcommand{\msc}{\operatorname{msc}}

\newcommand{\Alphabet}{\Sigma}

\newcommand{\MsgVals}{\ensuremath{\mathcal{V}}}

\newcommand{\CSM}[1]{\ensuremath{\{\!\!\{#1_\procA\}\!\!\}_{\procA \in \Procs}}}

\newcommand{\emptystring}{\varepsilon}

\newcommand{\set}[1]{\{#1\}}
\newcommand{\lang}{\mathcal{L}}

\newcommand{\interswaplang}{\mathcal{C}^{\interswap}}

\newcommand{\channels}{\ensuremath{\mathsf{Chan}}}
\newcommand{\channel}[2]{\ensuremath{\langle#1,#2\rangle}}

\newcommand{\expansion}{\operatorname{Exp}}
\newcommand{\trace}{\operatorname{trace}}

\newcommand{\semglobal}{\ensuremath{\mathsf{GAut}}}

\newcommand{\interswap}{\ensuremath{\sim}}

\def \ifempty#1{\def\temp{#1} \ifx\temp\empty }

\newcommand{\snd}[3]{\ifempty{#1} #2!#3 \else #1\triangleright#2!#3 \fi}
\newcommand{\rcv}[3]{\ifempty{#2} #1?#3 \else #2\triangleleft#1?#3 \fi}
\newcommand{\msgFromTo}[3]{#1\!\to\!#2\!:\!#3}

\newcommand{\pref}{\operatorname{pref}}
\newcommand{\preforder}{\ensuremath{\leq}}

\newcommand{\channelcompliant}{channel-compliant\xspace}

\newcommand{\Channelcompliant}{Channel-compliant\xspace}

\newcommand{\wproj}{{\ensuremath{\Downarrow}}}

\def\mmerge{\mathrel{\ThisStyle{\stretchrel*{\ooalign{\raise0.2\LMex\hbox{$\SavedStyle\sqcap$}\cr \raise-0.2\LMex\hbox{$\SavedStyle\sqcap$}}}{\sqcap}}}}

\newcommand{\qOPT}{\operatorname{qOPT}}

\newcommand{\mpstTOhmscc}{H}

\newcommand{\match}{\vdash\hspace{-4pt}\dashv}

\newcommand*\hypo[1]
{\tikz[baseline=(char.base)]{
            \node[shape=rectangle,draw,inner sep=1.5pt] (char) {#1};}}

\newcommand*\hyporounded[1]
{\tikz[baseline=(char.base)]{
            \node[shape=rectangle,rounded corners =0.15cm, draw,inner sep=1.5pt] (char) {#1};}}
\newcommand*\hypouse[1]{#1}

\newcommand{\union}{\cup}
\newcommand{\inters}{\cap}
\newcommand{\Union}{\bigcup}

\DeclarePairedDelimiter\card{\lvert}{\rvert}

 \providecommand{\Coloneqq}{\mathrel{\mathop{::}}=} \newcommand{\is}{\coloneq}

\newcommand{\from}{\colon}
\newcommand{\pto}{\rightharpoonup}
\newcommand{\inv}[1]{#1^{-1}}

\def\grammOr{\hspace{3pt}\mid\hspace{3pt}}
\def\grammIs{\Coloneqq}

\begingroup
\catcode`\|=\active \gdef\@grammar@bar{\catcode`\|=\active \def|{\grammOr}}
\endgroup

\newcommand{\gramm}[1]{\begingroup
  \def\is{\grammIs}\@grammar@bar #1\endgroup }

\newenvironment{grammar}{\begin{equation*}\def\is{& \grammIs }\@grammar@bar \aligned }
{\endaligned \end{equation*}\aftergroup\ignorespaces }

\newcommand{\qed}{$\square$}

\newcommand{\hole}{\hbox{-}}

\DeclarePairedDelimiterXPP\aenc[2]{\constr{a}}{(}{)}{_{#2}}{\strip@parens#1}
\DeclarePairedDelimiterXPP\pub[1]{\constr{p}}{(}{)}{}{\strip@parens#1}

\newcommand{\Parallel}{\@ifstar{\prod}{{\textstyle\prod}}}
\newcommand{\Alt}{\@ifstar{\sum}{{\textstyle\sum}}}
\newcommand{\Sum}{\Sigma}

 \usepackage{subcaption}
\usepackage{graphicx}

\usepackage[noend]{algpseudocode}
\usepackage{algorithm}

\definecolor{colorblind1}{RGB}{216, 27, 96} \definecolor{colorblind2}{RGB}{30, 136, 229} \definecolor{colorblind3}{RGB}{255, 193, 7} \definecolor{colorblind4}{RGB}{0, 77, 64} 

\usepackage[capitalise]{cleveref}
\usepackage{newunicodechar}
\newunicodechar{∃}{\ensuremath{\exists}}
\newunicodechar{∀}{\ensuremath{\forall}}
\newunicodechar{θ}{\ensuremath{\theta}}
\newunicodechar{τ}{\ensuremath{\tau}}
\newunicodechar{φ}{\ensuremath{\varphi}}
\newunicodechar{ξ}{\ensuremath{\xi}}
\newunicodechar{ζ}{\ensuremath{\zeta}}
\newunicodechar{ψ}{\ensuremath{\psi}}
\newunicodechar{π}{\ensuremath{\pi}}
\newunicodechar{α}{\ensuremath{\alpha}}
\newunicodechar{β}{\ensuremath{\beta}}
\newunicodechar{γ}{\ensuremath{\gamma}}
\newunicodechar{δ}{\ensuremath{\delta}}
\newunicodechar{ε}{\ensuremath{\varepsilon}}
\newunicodechar{κ}{\ensuremath{\kappa}}
\newunicodechar{λ}{\ensuremath{\lambda}}
\newunicodechar{μ}{\ensuremath{\mu}}
\newunicodechar{ρ}{\ensuremath{\rho}}
\newunicodechar{σ}{\ensuremath{\sigma}}
\newunicodechar{ω}{\ensuremath{\omega}}
\newunicodechar{Γ}{\ensuremath{\Gamma}}
\newunicodechar{Φ}{\ensuremath{\Phi}}
\newunicodechar{Δ}{\ensuremath{\Delta}}
\newunicodechar{Σ}{\ensuremath{\Sigma}}
\newunicodechar{Π}{\ensuremath{\Pi}}
\newunicodechar{∑}{\ensuremath{\Sigma}}
\newunicodechar{∏}{\ensuremath{\Pi}}
\newunicodechar{Θ}{\ensuremath{\Theta}}
\newunicodechar{Ω}{\ensuremath{\Omega}}
\newunicodechar{⇒}{\ensuremath{\Rightarrow}}
\newunicodechar{⇐}{\ensuremath{\Leftarrow}}
\newunicodechar{⇔}{\ensuremath{\Leftrightarrow}}
\newunicodechar{→}{\ensuremath{\rightarrow}}
\newunicodechar{←}{\ensuremath{\leftarrow}}
\newunicodechar{↔}{\ensuremath{\leftrightarrow}}
\newunicodechar{¬}{\ensuremath{\neg}}
\newunicodechar{∧}{\ensuremath{\land}}
\newunicodechar{∨}{\ensuremath{\lor}}
\newunicodechar{≠}{\ensuremath{\neq}}
\newunicodechar{≡}{\ensuremath{\equiv}}
\newunicodechar{∼}{\ensuremath{\sim}}
\newunicodechar{≈}{\ensuremath{\approx}}
\newunicodechar{≥}{\ensuremath{\geq}}
\newunicodechar{≤}{\ensuremath{\leq}}
\newunicodechar{≫}{\ensuremath{\gg}}
\newunicodechar{≪}{\ensuremath{\ll}}
\newunicodechar{∅}{\ensuremath{\emptyset}}
\newunicodechar{⊆}{\ensuremath{\subseteq}}
\newunicodechar{⊂}{\ensuremath{\subset}}
\newunicodechar{∩}{\ensuremath{\cap}}
\newunicodechar{⋂}{\ensuremath{\cap}}
\newunicodechar{∪}{\ensuremath{\cup}}
\newunicodechar{⋃}{\ensuremath{\cup}}
\newunicodechar{⊎}{\ensuremath{\uplus}}
\newunicodechar{∈}{\ensuremath{\in}}
\newunicodechar{∉}{\ensuremath{\not\in}}
\newunicodechar{⊤}{\ensuremath{\top}}
\newunicodechar{⊥}{\ensuremath{\bot}}
\newunicodechar{₀}{\ensuremath{_0}}
\newunicodechar{₁}{\ensuremath{_1}}
\newunicodechar{₂}{\ensuremath{_2}}
\newunicodechar{₃}{\ensuremath{_3}}
\newunicodechar{₄}{\ensuremath{_4}}
\newunicodechar{₅}{\ensuremath{_5}}
\newunicodechar{₆}{\ensuremath{_6}}
\newunicodechar{₇}{\ensuremath{_7}}
\newunicodechar{₈}{\ensuremath{_8}}
\newunicodechar{₉}{\ensuremath{_9}}
\newunicodechar{⁰}{\ensuremath{^0}}
\newunicodechar{¹}{\ensuremath{^1}}
\newunicodechar{²}{\ensuremath{^2}}
\newunicodechar{³}{\ensuremath{^3}}
\newunicodechar{⁴}{\ensuremath{^4}}
\newunicodechar{⁵}{\ensuremath{^5}}
\newunicodechar{⁶}{\ensuremath{^6}}
\newunicodechar{⁷}{\ensuremath{^7}}
\newunicodechar{⁸}{\ensuremath{^8}}
\newunicodechar{⁹}{\ensuremath{^9}}
\newunicodechar{𝔹}{\ensuremath{\mathbb{B}}}
\newunicodechar{ℝ}{\ensuremath{\mathbb{R}}}
\newunicodechar{ℕ}{\ensuremath{\mathbb{N}}}
\newunicodechar{ℂ}{\ensuremath{\mathbb{C}}}
\newunicodechar{ℚ}{\ensuremath{\mathbb{Q}}}
\newunicodechar{𝕋}{\ensuremath{\mathbb{T}}}
\newunicodechar{𝕏}{\ensuremath{\mathbb{X}}}
\newunicodechar{ℤ}{\ensuremath{\mathbb{Z}}}
\newunicodechar{✓}{\checkmark}
\newunicodechar{✗}{\ensuremath{\times}}
\newunicodechar{◊}{\ensuremath{\lozenge}}
\newunicodechar{□}{\ensuremath{\square}}
\newunicodechar{𝓐}{\ensuremath{\mathcal{A}}}
\newunicodechar{𝓑}{\ensuremath{\mathcal{B}}}
\newunicodechar{𝓒}{\ensuremath{\mathcal{C}}}
\newunicodechar{𝓓}{\ensuremath{\mathcal{D}}}
\newunicodechar{𝓔}{\ensuremath{\mathcal{E}}}
\newunicodechar{𝓕}{\ensuremath{\mathcal{F}}}
\newunicodechar{𝓖}{\ensuremath{\mathcal{G}}}
\newunicodechar{𝓗}{\ensuremath{\mathcal{H}}}
\newunicodechar{𝓘}{\ensuremath{\mathcal{I}}}
\newunicodechar{𝓙}{\ensuremath{\mathcal{J}}}
\newunicodechar{𝓚}{\ensuremath{\mathcal{K}}}
\newunicodechar{𝓛}{\ensuremath{\mathcal{L}}}
\newunicodechar{𝓜}{\ensuremath{\mathcal{M}}}
\newunicodechar{𝓝}{\ensuremath{\mathcal{N}}}
\newunicodechar{𝓞}{\ensuremath{\mathcal{O}}}
\newunicodechar{𝓟}{\ensuremath{\mathcal{P}}}
\newunicodechar{𝓠}{\ensuremath{\mathcal{Q}}}
\newunicodechar{𝓡}{\ensuremath{\mathcal{R}}}
\newunicodechar{𝓢}{\ensuremath{\mathcal{S}}}
\newunicodechar{𝓣}{\ensuremath{\mathcal{T}}}
\newunicodechar{𝓤}{\ensuremath{\mathcal{U}}}
\newunicodechar{𝓥}{\ensuremath{\mathcal{V}}}
\newunicodechar{𝓦}{\ensuremath{\mathcal{W}}}
\newunicodechar{𝓧}{\ensuremath{\mathcal{X}}}
\newunicodechar{𝓨}{\ensuremath{\mathcal{Y}}}
\newunicodechar{𝓩}{\ensuremath{\mathcal{Z}}}
\newunicodechar{…}{\ensuremath{\ldots}}
\newunicodechar{∗}{\ensuremath{\ast}}
\newunicodechar{⊢}{\ensuremath{\vdash}}
\newunicodechar{⊧}{\ensuremath{\models}}
\newunicodechar{′}{\ensuremath{'}}
\newunicodechar{″}{\ensuremath{''}}
\newunicodechar{‴}{\ensuremath{'''}}
\newunicodechar{∥}{\ensuremath{\|}}
\newunicodechar{⊕}{\ensuremath{\oplus}}
\newunicodechar{⁺}{\ensuremath{^+}}
\newunicodechar{⊇}{\ensuremath{\supseteq}}
\newunicodechar{∘}{\ensuremath{\circ}}
\newunicodechar{∙}{\ensuremath{\cdot}}
\newunicodechar{⋅}{\ensuremath{\cdot}}
\newunicodechar{≈}{\ensuremath{\approx}}
\newunicodechar{×}{\ensuremath{\times}}
\newunicodechar{∞}{\ensuremath{\infty}}
\newunicodechar{⊑}{\ensuremath{\sqsubseteq}}
 \usepackage{mathpartir}

\newtoggle{app:proof}
\toggletrue{app:proof}

\title{
Comparing Channel Restrictions of \\
Communicating State Machines, \\
High-level Message Sequence Charts, \\
and Multiparty Session Types
}

\def\titlerunning{Comparing Channel Restrictions of CSMs, HMSCs, and MSTs}
\def\authorrunning{Felix Stutz and Damien Zufferey}

\author{
Felix Stutz \qquad
Damien Zufferey \institute{MPI-SWS, Kaiserslautern, Germany}
\email{\{fstutz,zufferey\}@mpi-sws.org}
}

\begin{document}

\maketitle

\begin{abstract}
Communicating state machines provide a formal foundation for distributed computation.
Unfortunately, they are Turing-complete and, thus, challenging to analyse.
In this paper, we classify \mbox{restrictions} on channels which have been proposed to work around the undecidability of verification questions.
We compare half-duplex communication, existential $B$-boundedness, and $k$-synchroni\-sability. These restrictions do not prevent the communication channels from growing arbitrarily large but still restrict the power of the model.
Each restriction gives rise to a set of languages so, for every pair of \mbox{restrictions}, we check whether one subsumes the other or if they are \mbox{incomparable}.
We investigate their relationship in two different contexts:
first, 
the one of communicating state machines, and, 
second, 
the one of communication protocol specifications using high-level message \mbox{sequence} charts.
Surprisingly, these two contexts yield different conclusions.
In addition, we \mbox{integrate} multiparty session types, another approach to specify communication protocols, into our classification.
We show that multiparty session type languages are half-duplex, existentially $1$-bounded, and \mbox{$1$-synchronisable}.
To~show this result, we provide the first formal embedding of multiparty session types into high-level message sequence charts.
 \end{abstract}

\paragraph{Acknowledgements and Funding.}
The authors would like to thank Emanuele D'Osualdo, Georg \mbox{Zetzsche}
and the anonymous reviewers for their feedback and suggestions.
This research was funded in part by the Deutsche Forschungsgemeinschaft project 389792660-TRR 248.

\section{Introduction}
\label{sec:intro}

Communicating state machines (CSMs) are one of the foundational models of message-passing concurrency.
Unfortunately, the combination of multiple processes and unbounded FIFO channels yields a Turing-complete model of computation even when the processes are finite-state \cite{DBLP:journals/jacm/BrandZ83}.
The communication channels can be used as memory and, therefore, most verification questions for CSMs are not algorithmically solvable.
To regain decid\-ability, one needs to exploit properties of specific systems.
For instance, if all the runs of some communicating state machine use finite memory, it is possible to verify this system.
This restriction, known as universal boundedness~\cite{DBLP:journals/fuin/GenestKM07}, admits only systems with finitely many reachable states.

In this paper, we compare three channel restrictions which allow infinite state systems while making interesting verification questions decidable.
We compare half-duplex communication \cite{DBLP:journals/iandc/CeceF05}, existential $B$-boundedness \cite{DBLP:journals/fuin/GenestKM07}, and $k$-synchronisability \cite{DBLP:conf/cav/BouajjaniEJQ18,DBLP:conf/fossacs/GiustoLL20}.

\begin{figure}[t]
\subcaptionbox{
    Communicating state machine: one state machine for $\procA$ (top) and one for $\procB$ (bottom)
    \label{fig:intro-csm}
}[0.40\textwidth]
{
\centering
\includegraphics[width=0.32\textwidth]{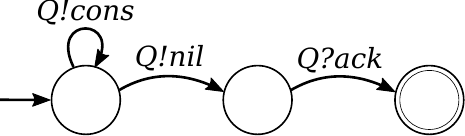}\\
\vspace{1.5ex}
\includegraphics[width=0.32\textwidth]{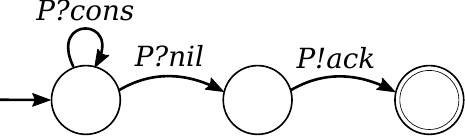}
}
\hfill
\subcaptionbox{
    High-level message sequence chart
    \label{fig:intro-hmsc}
}[0.23\textwidth]
{
\centering
\includegraphics[width=0.095\textwidth]{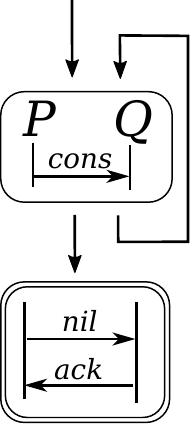}
}
\hfill
\subcaptionbox{
    Multiparty session type
    \label{fig:intro-mst}
}[0.35\textwidth]
{
    \centering
    \[
    μt. + \begin{cases}
    \msgFromTo{\procA}{\procB}{\mathit{cons}}.\,t \\
    \msgFromTo{\procA}{\procB}{\mathit{nil}}.\,\msgFromTo{\procB}{\procA}{\mathit{ack}}.\,0
    \end{cases}
    \]
\vspace{3ex}
}
\caption{Sending a list expressed in different formalisms.
    The left part is an implementation of the protocol specified in the middle and right parts.
}
\label{fig:intro}
\end{figure}

We explain all three restrictions with the CSM in \cref{fig:intro-csm}.
There, a process~$\procA$ sends a list, element by element, to a process~$\procB$.
After receiving the list's end, $\procB$ sends an acknowledgement back to~$\procA$.

\emph{Half-duplex communication} requires that, at all times, at least one of both channels between two processes is empty.
While $\procA$ sends the list, the channel can grow arbitrarily large.
However, $\procB$ always receives all the messages until $\mathit{nil}$ before replying.
When $\procB$ replies, the channel from $\procA$ to $\procB$ is empty.
Hence, the CSM is half-duplex.

\emph{Existential \mbox{$B$-boundedness}} means that, for every execution, we can reorder the sends and receptions such that the channels carry at most $B$ messages.
This CSM is existentially $1$-bounded.
Each reception is possible directly after the send.

\emph{$k$-synchronisability} requires that every execution can be reordered and split into phases where up to~$k$ messages are first sent and then received.
This CSM is $1$-synchronisable because every message can be received directly after it was sent.

The original definitions of channel restrictions are phrased in terms of executions of a CSM.
We present a characterisation for each restriction which only considers the generated language.
This also allows us to reason about languages specified or generated in different ways.
We consider languages given by protocol specifications and implementations.
For implementations, we consider \emph{CSM-definable languages}, i.e., languages which can be generated by a CSM.

\begin{figure}[t]
\begin{subfigure}{0.48\textwidth}
\centering
\resizebox{0.8\textwidth}{!}{
\begin{tikzpicture}
  \tikzset{venn circle/.style={draw,ellipse,minimum width=4cm,minimum height=3cm,opacity=0.5}}

  \node [venn circle = white] (A) at (0,0) {};
  \node [xshift=-1.0cm,yshift=-0.0cm] (AT) at (A) {$\exists B$-bounded};
  \node [yshift=-0.5cm] at (AT.south) {\color{colorblind2}{$\hypo{C3}$}};
  \node [venn circle = white] (B) at (60:2.0cm) {};
  \node [yshift=0.4cm] (BT) at (B) {half-duplex};
  \node [yshift=0.3cm] at (BT.north) {\color{colorblind3}{$\hypo{C7}$}};
  \node [venn circle = white] (C) at (0:2.0cm) {};
  \node [align=center, xshift=1.0cm,yshift=-0.0cm] (CT) at (C) {$k$-synch-\\ronisable};
  \node [yshift=-0.3cm] at (CT.south) {\color{colorblind1}{$\hypo{C5}$}};
  \node[left,xshift=-0.2cm,yshift=0.1cm] at (barycentric cs:A=1/2,B=1/2 ) {\color{colorblind2}{$\hypo{C2}$}};
  \node[below] at (barycentric cs:A=1/2,C=1/2 ) {\color{colorblind2}{$\hypo{C4}$}};
  \node[right,xshift=0.2cm,yshift=0.1cm] at (barycentric cs:B=1/2,C=1/2 ) {\color{colorblind2}{$\hypo{C6}$}};
  \node[below,yshift=0.2cm] at (barycentric cs:A=1/3,B=1/2,C=1/3 ){\color{colorblind2}{$\hypo{C1}$}};
  \node [draw, rectangle, minimum width=7.5cm, minimum height=5.8cm,yshift=0.5cm, rounded corners = 0.5cm] (Z) at (barycentric cs:A=1/3,B=1/3,C=1/3) {};
  \node [yshift=-0.5cm, xshift=1.4cm] (ZT) at (Z.north west) {\textbf{CSM-definable}};
  \node [yshift=-0.6cm,xshift=-0.3cm] at (ZT.south) {\color{colorblind3}{$\hypo{C8}$}};
\end{tikzpicture}
 }
\subcaption{CSM-definable Languages}
\label{fig:overview-csm}
\end{subfigure}
\hfill
\begin{subfigure}{0.48\textwidth}
\centering
\resizebox{0.8\textwidth}{!}{
\begin{tikzpicture}
  \node [draw, rectangle, rounded corners = 0.5cm, minimum width = 8.2cm, minimum height = 6.3cm, opacity=0.5] (Bbounded) at (0.1,-0.5) {};
  \node [yshift=-0.3cm, xshift=-1.4cm] (AT2) at (Bbounded.north east) {$\exists B$-bounded};
  \node [draw, rectangle, rounded corners = 0.5cm, minimum width = 7.5cm, minimum height = 5.0cm] (A) at (0,-0.9cm) {};
  \node [yshift=-0.5cm, xshift=1.6cm] (AT1) at (A.north west) {\textbf{HMSC-definable}};
\node[yshift=0.13cm, xshift=0.03cm, rounded corners = 0.15cm] (H1) at (A.north east) {\color{colorblind3}{$\hypo{H1}$}};

  \node [draw, rectangle, rounded corners = 0.3cm, minimum width = 4.5cm, minimum height = 3.8cm, opacity = 0.5] (B) at (-1.0cm,-1.2cm) {};
  \node [yshift=-0.4cm, xshift=1.5cm] (BT) at (B.north west) {$k$-synchronisable};
  \node[yshift=-0.2cm] (H2) at (BT.south) {\color{colorblind2}{$\hypo{H2}$}};
  \node[yshift=-0.2cm, xshift=3.7cm] (H3) at (BT.south) {\color{colorblind2}{$\hypo{H3}$}};

  \node [draw, rectangle, rounded corners = 0.3cm, minimum width = 6.5cm, minimum height = 2.3cm, opacity = 0.5] (C) at (0.3cm,-1.7cm) {};
  \node [yshift=-0.4cm, xshift=-1cm] (CT) at (C.north east) {half-duplex};
  \node[yshift=-0.8cm] (H4) at (CT.south) {\color{colorblind2}{$\hypo{H4}$}};

  \node [draw, rectangle, rounded corners = 0.3cm, minimum width = 2.7cm, minimum height = 1.6cm, opacity = 0.5] (D) at (-1.3cm,-1.7cm) {};
  \node [yshift=-0.4cm, xshift=0.8cm, align=right] (DT) at (D.north west) {$1$ sync.};
  \node[yshift=-0.3cm,xshift=2.5cm] (H5) at (DT.south) {\color{colorblind2}{$\hypo{H5}$}};
  \node[yshift=0cm,xshift=0.15cm, rounded corners = 0.15cm] (H6) at (D.north east) {\color{colorblind2}{$\hypo{H6}$}};
  \node[yshift=-0.2cm] (H7) at (DT.south) {\color{colorblind2}{$\hypo{H7}$}};

  \node [draw, rectangle, rounded corners = 0.3cm, minimum width = 1.1cm, minimum height = 1.3cm, opacity = 0.5] (E) at (-0.6cm,-1.7cm) {};
  \node [yshift=-0.55cm, xshift=0.0cm, align=center] (ET) at (E.north) {\textbf{MST}\\\textbf{-def.}};
  \node[yshift=-0.6cm,xshift=-0.77cm, rounded corners = 0.15cm] (H8) at (ET.south) {\color{colorblind2}{$\hypo{H8}$}};
\end{tikzpicture}
 }
\subcaption{HMSC-definable Languages}
\label{fig:overview-hmsc}
\end{subfigure}\caption[
]{
    Comparing half-duplex, existential $B$-bounded, and $k$-synchronisable systems.
    The \textcolor{colorblind3}{results} are known results,
    \textcolor{colorblind2}{results} are new, and
    the \textcolor{colorblind1}{result} disproves an existing result.
    Hypotheses with
    \hyporounded{rounded}
    corners indicate inclusions while
    \hypo{pointed}
    corners indicate incomparability results.
}
\label{fig:overview-relations}
\end{figure}
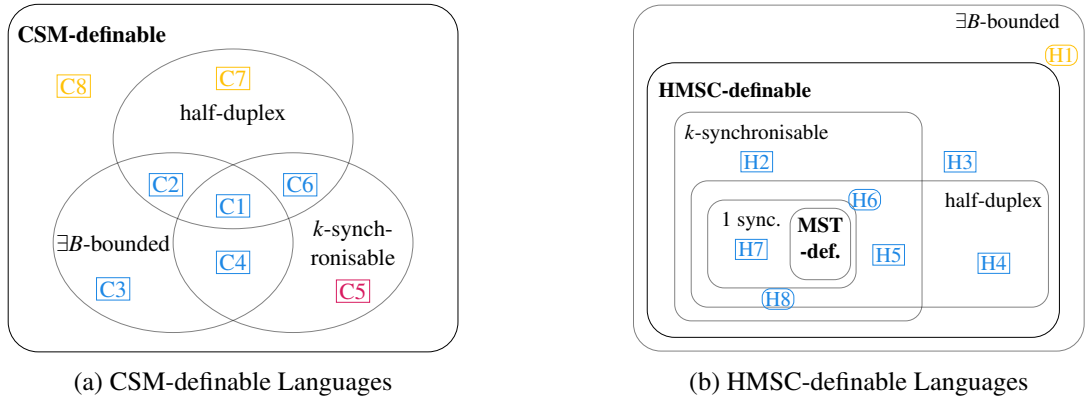

Interestingly, for CSMs, these channel restrictions have not yet been compared thoroughly.
In this paper, we close this gap and provide a classification of channel restrictions for CSM-definable languages.
For instance, this answers a question for the FIFO point-to-point setting which has been posed for the mailbox setting by Bouajjani et al.~\cite{DBLP:conf/cav/BouajjaniEJQ18} as we prove that existential \mbox{$B$-boundedness} and \mbox{$k$-synchronisability} are incomparable for CSM-definable languages.
Overall, we give examples for every possible intersection and, thus, prove that none of the restrictions subsumes another one in this context. Our results for CSM-definable languages are summarised in \cref{fig:overview-csm}.
In fact, we disprove one of the three known results from the literature \cite[Thm.~7.1]{DBLP:conf/cav/LangeY19} which has been cited recently as part of a summary~\cite[Prop.~41]{DBLP:conf/concur/BolligGFLLS21}.
This indicates that, despite their simplicity, these definitions hide some subtleties.
Our classification provides a careful treatment --- giving minimal examples for the sake of~understandability.

Such a classification is interesting as focusing on languages or systems adhering to one of the channel restrictions can be key for solving verification problems algorithmically.
For instance, control-state reachability and model checking LCPDL (propositional dynamic logic with loop and converse) formulas are decidable for $k$-synchronisable systems \cite{DBLP:conf/fossacs/GiustoLL20,DBLP:conf/concur/BolligGFLLS21}.
Later, we highlight the impact of channel restrictions on verification questions and whether one can check if a system adheres to a restriction.

\smallskip\noindent\textbf{Protocol Specifications.}
Instead of considering arbitrary CSMs, it is possible to start with a global description written in a dedicated protocol specification formalism such as
    High-level Message \mbox{Sequence} Charts (HMSCs)~\cite{DBLP:conf/concur/AlurY99,DBLP:conf/ac/GenestMP03}, 
    Multiparty Session Types (MSTs)~\cite{DBLP:conf/popl/HondaYC08,DBLP:journals/jacm/HondaYC16}, or
    Choreography Automata (CA)~\cite{DBLP:conf/coordination/BarbaneraLT20}.
A~protocol is a global specification of all the processes' actions together while an implementation only gives the local actions of each process.
\cref{fig:intro} shows, along the CSM, two protocol specifications.
The key difference between a protocol specification and an implementation is that the protocol specification \mbox{explicitly} connects a send event to the corresponding receive event.
In the HMSC~(Fig.~\ref{fig:intro-hmsc}), the arrows connect sends to receptions.
The MST\footnote{
\label{footnote:mst-global-type}
We actually present a global type in an MST framework here but only use the term after its formal introduction in \cref{sec:mst}}
~(Fig.~\ref{fig:intro-mst})
specifies communication by ${\msgFromTo{\mathit{sender}}{\mathit{receiver}}{\mathit{message}}}$.
The CSM~(Fig.~\ref{fig:intro-csm}) does not specify this connection upfront~and it may not exist.
This makes CSMs strictly more general than protocols.
For instance, an incorrect implementation of the protocol could have $\procA$ terminate before receiving the acknowledgement.

The CSM, HMSC, and MST all have the same language. Thus, our observations on channel restrictions also hold for the HMSC and the MST.
We also say that the CSM \emph{implements} the protocol specified by the HMSC (or the MST) as they accept the same language and the CSM is deadlock free.
In general, there are several approaches to obtain a CSM which implements a protocol (if one exists).
For instance, a protocol specification can be projected on to each process. In this paper, we do not consider this \mbox{problem}.
A protocol specification gives rise to a language, i.e., the protocol.
We only need the protocol as our definitions for channel restrictions apply to languages, e.g.,
\emph{HMSC-} and \emph{MST-definable languages}.

For protocols, the classification of channel restrictions was less studied than for CSMs.
\cref{fig:overview-hmsc} summarises our results.
It was only known that each HMSC-definable language is existentially $B$-bounded for some~$B$~\cite{DBLP:journals/fuin/GenestKM07}.
Surprisingly, the classification changes in the context of protocols.
For restrictions which differ ($\hypo{H2}$ to $\hypo{H5}$, and $\hypo{H7}$), we give distinguishing examples.
When one restriction subsumes another one ($\hyporounded{H1}$, $\hyporounded{H6}$, and $\hyporounded{H8}$), we prove it.
For instance, $\hyporounded{H6}$ proves that $1$-synchronisability entails half-duplex communication while $\hypo{H5}$ is an example which is half-duplex, existentially $B$-bounded, $k$-synchronisable but not $1$-synchronisable.

\smallskip\noindent\textit{Embedding MSTs into HMSCs.}
In addition to our results about CSM- and HMSC-definable languages, we provide the first formal embedding from MSTs into \mbox{HMSCs}.
The contribution is two-fold.
First, we situate MSTs in the picture of common channel restrictions and prove that languages specified by multiparty session types are half-duplex, existentially $1$-bounded, and $1$-synchronisable.
This sheds a new light on why MSTs are effectively analysable.
Second, we did recently show that using insights from the domain of HMSCs in the domain of MSTs is a promising research direction as we made the effective MST verification techniques applicable to patterns from distributed computing~\cite{DBLP:conf/concur/MajumdarMSZ21}.
Hence, our formal embedding can act as a crucial building block for further advances which are facilitated by insights from both domains.

\smallskip\noindent\textbf{Contributions.}
In this paper, we make three main contributions.
(1) We provide an exhaustive classification of channel restrictions for CSM-definable languages.
In this process, we disprove a recent result from the literature. (2) We provide an exhaustive classification of channel restrictions for HMSC- and MST-definable languages.
(3) We give the first formal embedding of MSTs into HMSCs.

\smallskip\noindent\textbf{Outline.}
After providing some preliminary definitions in \cref{sec:prelim}, we define the channel restrictions formally in \cref{sec:channel-restrictions} and summarise their impact on the decidability of verification questions.
Subsequently, we establish our results on HMSCs (\cref{sec:hmsc}), MSTs (\cref{sec:mst}), and CSMs (\cref{sec:implementing-protocols}).
We discuss related work in \cref{sec:related}.
 \section{Preliminaries}
\label{sec:prelim}

\smallskip\noindent\textbf{Finite and Infinite Words.}
For an alphabet $\Sigma$, the set of finite words over $\Sigma$ is denoted by~$\Sigma^*$,
the set of infinite words by $\Sigma^\omega$, while we write $\Sigma^\infty = \Sigma^* \cup \Sigma^\omega$ for their union.
For two strings $u\in\Sigma^*$ and $v\in\Sigma^\infty$, $u$~is said to be a \emph{prefix} of $v$, denoted by $u \leq v$,
if there is some $w\in\Sigma^\infty$ such that $u \cdot w = v$.
For two alphabets $\Sigma$ and $\Delta$ with $\Delta \subseteq \Sigma$,
the \emph{projection} of $w \in\Sigma^\infty$ on to~$\Delta$,
denoted by $w\wproj_\Delta$, is the
word which is obtained by omitting every letter in $w$ that does not belong to~$\Delta$.

\smallskip\noindent\textbf{Message Alphabet.}
$\Procs$ is a finite set of processes, ranged over by $\procA, \procB, \procC, \ldots$, and $\MsgVals$ a finite set of messages.
For a process $P$, we define the alphabet
    $Σ_{\procA} = \set{ \snd{\procA}{\procB}{\val}, \rcv{\procB}{\procA}{\val} \mid \procB \in \Procs,\; \val \in \MsgVals }$ of events.
The event $\snd{\procA}{\procB}{\val}$ denotes process $\procA$ sending a message $\val$ to $\procB$,
and $\rcv{\procB}{\procA}{\val}$ denotes process $\procA$ receiving a message $\val$ from $\procB$.
Note that the process performing the action is always the first one, e.g., the receiver~$\procA$ in $\rcv{\procB}{\procA}{\val}$.
The alphabet $\Alphabet = \Union_{\procA \in \Procs} \Alphabet_{\procA}$ denotes all send and receive events while
$Σ_{\mathit{sync}} = \set{ \msgFromTo{\procA}{\procB}{\val} \mid \procA,\procB ∈ \Procs \text{ and } \val ∈ \MsgVals}$
is the set where sending and receiving a message is specified at the same time.
We fix $\Procs$, $\MsgVals$, $\Sigma$, and $\Sigma_{\mathit{sync}}$ in the rest of the paper.
We write $w\wproj_{\snd{\procA}{\procB}{\_}}$ to select all send events in $w$ where $\procA$~sends a message to $\procB$ and $\MsgVals(w)$ to project the send and receive events to their message~values.

\smallskip\noindent\textbf{Distributed Executions.}
We use these specialised alphabets to model specifications in which multiple distributed processes communicate by exchanging messages.
Furthermore, these executions cannot be any word but need to comply with conditions that correspond to the asynchronous communication over reliable FIFO channels.
We call such words \channelcompliant.

\begin{definition}[\cite{DBLP:conf/concur/MajumdarMSZ21}]A \emph{protocol} is a set of complete \channelcompliant words where:
\mbox{}
\begin{enumerate}
\item \emph{\Channelcompliant:}
    A word $w\in \Sigma^\infty$ is \channelcompliant if messages are received after they are sent and, between two processes,
    the reception order is the same as the send order.
    Formally, for each prefix $w'$ of $w$, we require
    $\MsgVals(w'\wproj_{\rcv{\procA}{\procB}{\_}})$ to be a prefix of
    $\MsgVals(w'\wproj_{\snd{\procA}{\procB}{\_}})$, for every $\procA,\procB \in \Procs$.
\item \emph{Complete:}
    A \channelcompliant word $w\in\Sigma^\infty$ is \emph{complete} if it is infinite or the send and receive events match:
    if $w \in \Sigma^*$, then $\MsgVals(w\wproj_{\snd{\procA}{\procB}{\_}}) = \MsgVals(w\wproj_{\rcv{\procA}{\procB}{\_}})$ for every $\procA,\procB \in \Procs$.
\end{enumerate}
\end{definition}

To pinpoint the corresponding send and receive events, we define a notion of \emph{matching}.

\begin{definition}[Matching Sends and Receptions]
In a word $w = e_1 \ldots \in \Alphabet^\infty$,
a send event $e_i = \snd{\procA}{\procB}{\val}$ is \emph{matched} by a receive event $e_j = \rcv{\procA}{\procB}{\val}$,
denoted by $e_i \match e_j$, if $i < j$ and
$\MsgVals((e_1 \ldots e_i) \wproj_{\snd{\procA}{\procB}{\_}})$
=
$\MsgVals((e_1 \ldots e_j) \wproj_{\rcv{\procA}{\procB}{\_}})$.
A~send event $e_i$ is \emph{unmatched} if there is no such receive event~$e_j$.
\end{definition}

If a sequence of events is \channelcompliant, it is trivial that for each channel between two processes, either all send events are matched or there is an index from which all send events are unmatched.

In this paper, we consider protocols that can be specified with high-level messages sequence charts.
We define \emph{prefix message sequence charts} to allow unmatched send events, inspired by the work of Genest et al.~\cite[Def.~3.1]{DBLP:journals/fuin/GenestKM07}.
The definition of a (prefix) MSC can look intimidating.
In \cref{fig:bmsc-color}, we show pictorially what each component corresponds to.

\begin{definition}[(Prefix) Message Sequence Charts]
\label{def:msc}
A \emph{prefix message sequence chart} is a $5$-tuple
\linebreak $M = (\textcolor{colorblind1}{\eventnodes},\textcolor{colorblind2}{p},\textcolor{colorblind3}{f},\textcolor{colorblind4}{l},(\leq_\procA)_{\procA \in \Procs })$
where

\noindent
\begin{minipage}{0.66\textwidth}
\begin{itemize}
\vspace{0.5ex}
\item $\textcolor{colorblind1}{\eventnodes}$ is a set of send $(\SndEvs)$ and receive $(\RcvEvs)$ event nodes $(\eventnodes = \SndEvs ⊎ \RcvEvs)$,\item $\textcolor{colorblind2}{p} \from \eventnodes \to \Procs$ maps each event node to the process acting on it,
\item $\textcolor{colorblind3}{f} \from \SndEvs \pto \RcvEvs$ is an injective partial function linking \\ corresponding send and receive event nodes,
\item $\textcolor{colorblind4}{l} \from \eventnodes \to Σ$ labels every event node with an event, and
\item $(\leq_\procA)_{\procA \in \Procs }$ is a family of total orders for the\\
event nodes of each process: $\leq_\procA \; \subseteq \; \inv{p}(\procA) \times \inv{p}(\procA)$.
\end{itemize}
\end{minipage}\begin{minipage}{0.05\textwidth}
\phantom{s}
\end{minipage}\begin{minipage}{0.28\textwidth}
\begin{center}
    \includegraphics[width=0.83\textwidth]{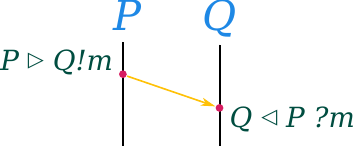}
  \end{center}
\captionof{figure}{Highlighting the elements of a (prefix) MSC: $(\textcolor{colorblind1}{\eventnodes},\textcolor{colorblind2}{p},\textcolor{colorblind3}{f},\textcolor{colorblind4}{l},(\leq_\procA)_{\procA \in \Procs })$}
  \label{fig:bmsc-color}
\end{minipage}
\vspace{1ex}

\noindent A prefix MSC $M$ induces a partial order $\leq_M$ on $\eventnodes$ that is defined co-inductively\footnote{
Note that we cannot use the standard reflexive and transitive closure since we consider infinite sequences of events.
Co-induction lifts the reflexive, transitive closure of the union of the send-receive relation and all process orders, i.e.,
   $
(
        \set{(s, f(s)) \mid s \in \SndEvs}
        \; \cup \;
        \bigcup_{\procA\in\Procs} \leq_\procA
)^*
   $,
to infinite sets of event nodes.
}\emph{:}
{
\normalsize
\begin{mathpar}

\inferrule*[right=proc]{\label{constr:proc}
e \leq_\procA e'
}{
e \leq_M e'
}

\inferrule*[right=snd-rcv]{\label{constr:snd-rcv}
s \in \SndEvs
}{
s \leq_M f(s)
}

\inferrule*[right=refl]{\label{constr:refl}
}{
e \leq_M e
}

\inferrule*[right=trans]{\label{constr:trans}
e \leq_M e' \\
e' \leq_M e''
}{
e \leq_M e''
}
\end{mathpar}
}
The labelling function $l$ respects the function $f$ between $\SndEvs$ and $\RcvEvs$:
for every pair of event nodes $e,e' \in \eventnodes$ with $f(e) = e'$, we have
$l(e) = \snd{p(e)}{p(e')}{\val}$ and $l(e') = \rcv{p(e)}{p(e')}{\val}$ for some $\val \in \MsgVals$ and for every $e$ where $f(e)$ is undefined, we have $l(e) = \snd{p(e)}{\procA}{\val}$ for some $\procA \neq p(e)$ according to its destination.
\end{definition}

We say that $M$ is \emph{degenerate} if there is some $\procA$ and $\procB$ such that there are $e_1, e_2 \in \inv{p}(\procA)$ with $e_1 \neq e_2 $, $l(e_1) = l(e_2)$, $e_1 \leq_\procA e_2$ and $f(e_2) \leq_\procB f(e_1)$.
We say that $M$ \emph{respects FIFO order} if
$M$ is not degenerate and
for every pair of processes $\procA$, $\procB$, and for every two event nodes $e_1 \leq_M e_2$ with $l(e_i) = \snd{\procA}{\procB}{\_}$\phantom{x}for $i \in \set{1,2}$, it holds that $f(e_2)$ is undefined if $f(e_1)$ is undefined as well as that
it holds that $\MsgVals(w_\procA) = \MsgVals(f(w_\procA))$ where $w_\procA$ is the (unique) linearisation of $\inv{p}(\procA)$.

In this paper, we do only consider prefix message sequence charts that respect FIFO order.

If $f$ is total, we omit the term \emph{prefix} and call $M$ a \emph{message sequence chart (MSC)}.
If $\eventnodes$ is finite for an MSC $M$, we call $M$ a \emph{basic MSC (BMSC)}.
We denote the set of BMSCs by $\BMSCs$.
When $M$ is clear from context, we simply write $\leq$ instead of~$\leq_M$.
For a prefix MSC~$M$, the language $\lang(M)$ contains a sequence $l(w)$
for each linearisation $w$ of $\eventnodes$ compatible with~$\leq_M$.
When unambiguous, we may refer to event nodes or sequences thereof by their (event) labels or omit the label function $l$.

A prefix MSC, in contrast to an MSC, allows send event nodes for any channel to be unmatched from some point on.
The concatenation $M_1 \cdot M_2$, or simply $M_1 M_2$, of an MSC $M_1$ and a prefix MSC $M_2$ is defined as expected (see \cref{def:concatenation-msc} in \cref{app:mscs} for the formal definition).
The concatenation requires that, for any individual process, all event nodes in $M_1$ happen before the
event nodes in $M_2$.
However, the induced partial order on $\eventnodes$ may permit linearisations in which an event node
from $M_2$ of one process occurs before an event node from $M_1$ of another process.

For every \channelcompliant word $w$, one can construct a unique prefix MSC $M$ such that $w$ is a linearisation of $M$.

\begin{lemma}[$\msc(\hole)$ (\cite{DBLP:journals/fuin/GenestKM07}, Section 3.1)]
\label{lm:msc-function}
Let $w \in \Alphabet^\infty$ be a \channelcompliant word.
Then, there is unique prefix MSC, denoted by $\msc(w)$,
such that $w$ is a linearisation of~$\msc(w)$.
In case the above conditions are not satisfied, $\msc(w)$ is undefined.
\end{lemma}

All sequences of events we consider in this work are \channelcompliant.
For sequences from MSCs (considered in \cref{sec:hmsc}), this trivially holds, while for sequences from execution prefixes of CSMs (considered in~\cref{sec:implementing-protocols}), \cref{lm:execution-prefix-and-channel-content} in \cref{app:half-duplex} ensures this.
 \section{Channel Restrictions}
\label{sec:channel-restrictions}

In this section, we present different channel restrictions and their implications on decidability of interesting verification questions.
Their application is discussed subsequently:
for HMSCs in \cref{sec:channel-hmsc},
for MSTs in \cref{sec:channeluse-mst},
and for 
CSMs in \cref{sec:understanding-channel-use}.

\subsection{Definitions}

\subsubsection{Half-duplex Communication}
C{\'{e}}c{\'{e}} and Finkel \cite[Def.~8]{DBLP:journals/iandc/CeceF05} introduced the restriction of half-duplex communication which intuitively requires that, for any two processes $\procA$ and $\procB$, the channel from $\procA$ to $\procB$ is empty before $\procB$ sends a message to $\procA$.
We define the restriction of half-duplex on sequences of events and show that it is equivalent to the original definition in \cref{app:half-duplex}.

\begin{definition}[Half-duplex]
\label{def:half-duplex}
A sequence of events $w$ is called \emph{half-duplex} if for every prefix $w'$ of $w$ and pair of processes $\procA$ and $\procB$,
one of the following holds:
$\MsgVals(w' \wproj_{\snd{\procA}{\procB}{\_}}) =
\MsgVals(w' \wproj_{\rcv{\procA}{\procB}{\_}})$ or
$\MsgVals(w' \wproj_{\snd{\procB}{\procA}{\_}}) =
\MsgVals(w' \wproj_{\rcv{\procB}{\procA}{\_}})$.
A~language $L \subseteq \Alphabet^\infty$ is half-duplex if every word $w \in L$ is.
\end{definition}

\subsubsection{Existential $B$-boundedness}
While the previous property restricts the channel for at least one direction to be empty, one can also bound the size of channels and consider linearisations that are possible adhering to such bounds.
On the one hand, one can consider a universal bound that applies for every linearisation.
However, this yields finite-state systems \cite{DBLP:journals/fuin/GenestKM07} and disallows very simple protocols, e.g., the example in \cref{fig:intro}.
On the other hand, one can consider an existential bound on the channels which solely asks that there is one linearisation of the distributed execution for which the channels are bounded.
This allows infinite-state systems and admits the earlier example.

\begin{definition}[$B$-bounded \cite{DBLP:journals/fuin/GenestKM07}]
Let $B \in \Nat$ be a natural number.
A word $w$ is \emph{$B$-bounded} if for every prefix $w'$ of $w$ and pair of processes $\procA$ and $\procB$, it holds that
$\card{w' \wproj_{\snd{\procA}{\procB}{\_}}} -
\card{w' \wproj_{\rcv{\procA}{\procB}{\_}}} \leq B$.
\end{definition}

\begin{definition}[Existentially $B$-bounded \cite{DBLP:journals/fuin/GenestKM07}]
Let $B \in \Nat$.
A prefix MSC~$M$ is \emph{existentially $B$-bounded} if there is a $B$-bounded linearisation for $M$.
A sequence of events $w$ is \emph{existentially $B$-bounded} if $\msc(w)$ is defined and \emph{existentially $B$-bounded}.
A language $L$ is \emph{existentially $B$-bounded} if every word $w
\in L$ is.
We may use \emph{not existentially bounded} as abbreviation for not existentially $B$-bounded for any $B$.

\end{definition}

\subsubsection{$k$-synchronisability}
The restriction of $k$-synchronisability was introduced for mailbox communication~\cite{DBLP:conf/cav/BouajjaniEJQ18} and later refined and adapted to the point-to-point setting \cite{DBLP:conf/fossacs/GiustoLL20}.
We define $k$-synchronisability following definitions by Giusto et al.~\cite[Defs.\ 6 and~7]{DBLP:conf/fossacs/GiustoLL20}.
The definition of $k$-synchronisability builds upon the notion when a prefix MSC is $k$-synchronous.
Its first condition requires that there is some linearisation of the prefix MSC while its second condition requires causal delivery to hold.
In contrast to the mailbox setting, the first condition always entails the second condition for the point-to-point setting.

\smallskip\noindent\textbf{Point-to-point Communication implies Causal Delivery.}
We first adapt the definition of causal delivery \cite[Def.~4]{DBLP:conf/fossacs/GiustoLL20} for point-to-point FIFO channels \cite[Section 6]{DBLP:conf/fossacs/GiustoLL20}.
Unfortunately, this discussion leaves room for interpreting what causal delivery exactly is for point-to-point systems.
Based on the description that a process $\procA$ can receive messages from two distinct processes $\procB$ and $\procC$ in any order, regardless of the dependency between the corresponding send events, we decided to literally adapt the definition of causal delivery as follows.

\begin{definition}[Causal delivery]
Let $M = (\eventnodes, p, f, l, (\leq_\procA)_{\procA \in \Procs})$ be an MSC.
We say that $M$ satisfies \emph{causal delivery} if there is a linearisation $w = e_1 \ldots$ of $\eventnodes$ such that for any two events $e_i \leq_M e_j$ with $e_i = \snd{\procA}{\procB}{\_}$ and $e_j = \snd{\procA}{\procB}{\_}$, either $e_j$ is unmatched in $w$ or there are $e_{i'} \leq_M e_{j'}$ such that $e_i \match e_{i'}$ and $e_j \match e_{j'}$ in~$w$.
\end{definition}

We show that $\msc(w)$ for every $w$ (if defined) satisfies causal delivery (as proven in \cref{app:causal-delivery-ex-sync}).

\begin{lemma}
\label{lm:msc-sat-causal-delivery}
Let $w \in \Alphabet^\infty$ such that $\msc(w)$ is defined.
Then, $\msc(w)$ satisfies causal delivery.
\end{lemma}

In combination with the fact that, given a linearisation $w$ of a prefix MSC~$M$, $\msc(w)$ is isomorphic to $M$, this yields that causal delivery is satisfied if there is a linearisation.

\begin{corollary}
 Every prefix MSC with a linearisation satisfies causal delivery.
\end{corollary}

With this, we can simplify the definition by omitting the second condition without changing its meaning.
In addition, we extend it to apply for MSCs with infinite sets of event nodes.

\begin{definition}[$k$-synchronous and $k$-syn\-chro\-nis\-able]
\label{def:ex-k-synchronous}
\label{def:ex-k-synchronisability}
Let $k \in \Nat$ be a positive natural number.
We say that a prefix MSC $M = (\eventnodes, p, f, l,
(\leq_\procA)_{\procA \in \Procs })$ is \emph{$k$-synchronous} if
\begin{enumerate}
 \item there is a linearisation of event nodes $w$ compliant with $\leq_M$ which can be split into a sequence of \emph{$k$-exchanges} (also called \emph{message exchange} if $k$ not given or clear from context), i.e.,
 $w = w_1 \ldots$ such that for all $i$, it holds that $l(w_i) \in S^{\leq k} \cdot R^{\leq k}$; and
\item for all $e$, $e'$ in $w$ such that $e \match e'$, there is some $i$ with $e$, $e'$ in $w_i$.\footnote{
 This is equivalent to the following:
 for all $e$ and $f(e)$ in $w$, there is some $i$ with $e$, $f(e)$ in $w_i$.
 }\end{enumerate}
A linearisation $w$ is $k$-synchronisable\footnote{
One could distinguish between universal and existential $k$-synchronisability, i.e., to distinguish the existence of a $k$-synchronisable linearisation rather than all linearisations being $k$-synchronisable.
However, the universal version does not make much sense in practice.
Thus, we omit the term existential.
}
if $\msc(w)$ is $k$-synchronous.
A language~$L$ is $k$-synchronisable if every word $w \in L$ is.
We may use \emph{not synchronisable} as abbreviation for not $k$-synchronisable for any $k$.
\end{definition}

\subsection{Algorithmic Verification and Channel Restrictions}
\label{sec:alg-verification}

It is important to note that we use the term \emph{restriction} as a property of a system which occurs naturally and not something that is imposed on its semantics.
However, both have a tight connection:
a system \emph{naturally} satisfies a restriction if \emph{imposing} the restriction does not change its possible behaviours.
If this is the case, one can exploit this for algorithmic verification and only check behaviours that satisfy the restriction without harming correctness.

For each channel restriction, we recall known results about checking membership and which verification problems become decidable.

\smallskip\noindent\textbf{Half-duplex Communication.}
For CSMs with two processes, membership is decidable \cite[Thm.~31]{DBLP:journals/iandc/CeceF05}.
The set of reachable configurations is computable in polynomial time which renders many verification questions like the \emph{unspecified reception problem} decidable (see \cite[Thm.~16]{DBLP:journals/iandc/CeceF05} for a detailed list of verification problems) while
model checking PLTL or CTL is still undecidable.
Half-duplex CSMs with more than two processes are Turing-powerful \cite[Thm.~38]{DBLP:journals/iandc/CeceF05} so
verification becomes undecidable and checking membership is of little interest.

\smallskip\noindent\textbf{Existential $B$-boundedness.}
For CSMs, membership is undecidable, unless CSMs are known to be deadlock free and $B$ is given~\cite[Fig.~3]{DBLP:journals/fuin/GenestKM07}.
For protocols, we will see that they are always existentially $B$-bounded for some $B$ and thus a correct implementation of a protocol also is.
It is quite straightforward that control-state reachability is decidable but not typically studied for these systems~\cite{DBLP:conf/concur/BolligGFLLS21}.
Intuitively, it can be solved by exhaustively enumerating the reachability graph of the CSM while pruning configurations exceeding the bound~$B$.
For HMSCs, model checking is undecidable for LTL \cite[Thm.~3]{DBLP:conf/concur/AlurY99} and decidable for MSO \cite[Thm.~1]{DBLP:conf/icalp/Madhusudan01}.

\smallskip\noindent\textbf{$k$-synchronisability.}
For CSMs, membership for a given $k$ is decidable in EXPTIME~\cite[Rem.~30]{DBLP:conf/concur/BolligGFLLS21}, originally shown decidable by Di Giusto et al.~\cite{DBLP:conf/fossacs/GiustoLL20}, while it is undecidable if $k$ is not given~\cite[Thm.~22]{DBLP:conf/concur/BolligGFLLS21}.
For HMSCs, both questions are decidable in polynomial time, while we show that MSTs are always $1$-synchronisable.
Model checking for $k$-synchronisable systems is decidable and in EXPTIME when formulas are represented in LCPDL.
This follows from combining that such systems have bounded (special) tree-width \cite[Prop.~28]{DBLP:conf/concur/BolligGFLLS21} and results by Bollig and Finkel~\cite{DBLP:journals/corr/abs-1904-06942}.
Control-state reachability was shown to be decidable for \mbox{$k$-synchronisable} systems \cite[Thm.~6]{DBLP:conf/fossacs/GiustoLL20}.
 \section{High-level Message Sequence Charts}
\label{sec:hmsc}

Message sequence charts have been used as compact representation for executions of CSMs.
The (prefix) message sequence charts obtained from different executions of CSMs can be analysed to determine which channel restriction is satisfied \cite{DBLP:journals/fuin/GenestKM07,DBLP:conf/fossacs/GiustoLL20}.
In addition, we do also use message sequence charts as building blocks for high-level message sequence charts \cite{DBLP:conf/sdl/MauwR97,z120-standard} which specify protocols.

We define these following the presentation by Alur et al.~\cite{DBLP:journals/tse/AlurEY03,DBLP:journals/tcs/AlurEY05}.
A~BMSC corresponds to ``straight line code'' in which each process follows a single sequence of event nodes.
A \emph{high-level message sequence chart} (HMSC) adds a regular control structure (branching and loops).

\begin{definition}[High-Level Message Sequence Charts]
A \emph{high-level message sequence chart} (HMSC) is a structure $(V, \edges, \vertexA^I\negmedspace, V^T\negmedspace\!, \mu)$ where
$V$ is a finite set of vertices,
$\edges \subseteq V \times V$ is a set of directed edges,
$\vertexA^I \in V$ is an initial vertex,
$V^T \subseteq V$ is a set of terminal vertices, and
$\mu: V \to \BMSCs$ is a function mapping every vertex to a BMSC.\end{definition}

To obtain the language of an HMSC, we start with initial paths through the HMSC.
As usual, we are interested only in maximal paths, i.e., either infinite or ending in a terminal vertex.
We can expand each such path into a sequence of BMSCs, concatenate this sequence of BMSCs, and take the language of the resulting MSC.
The language of the HMSC is the union of the languages of the MSCs generated by all its initial paths -- see \cref{app:semantics-hmsc} for the formal definition.
For simplicity, we assume every vertex in an HMSC is reachable from the initial vertex and every initial non-maximal path can be completed to a maximal~one.

\begin{example}
\Cref{fig:intro-hmsc} shows an HMSC composed of two BMSCs.
MSCs of the HMSC are obtained by following the control structure and concatenating the corresponding BMSCs.
The language contains all linearisations of these MSCs.
\end{example}

\subsection{Channel Restrictions of HMSCs}
\label{sec:channel-hmsc}

We say that an HMSC is half-duplex, existentially $B$-bounded or \mbox{$k$-synchronisable} respectively if its language~is.
It is straightforward that checking an HMSC for $k$-synchronisability amounts to checking its BMSCs.

\begin{proposition}
\label{cor:k-sync-hmsc}
An HMSC $H$ is $k$-synchronisable iff all BMSCs of $H$ are $k$-synchronous.
\end{proposition}

For the presentation of our results, we follow the numbering laid out in \cref{fig:overview-hmsc}.
Note that any BMSC can always be turned into a HMSC with a single initial and terminal vertex.
Therefore, it is trivial that all BMSC examples also apply to HMSCs.

\begin{lemma}[\cite{DBLP:journals/fuin/GenestKM07}, Prop.~3.1]
\hypouse{\emph{$\hyporounded{H1}$:}}
\label{lm:HMSC-ex-B-bounded}
Any HMSC $H$ is existentially $B$-bounded for some~$B$.
\end{lemma}
We prove this result in a slightly different way in \cref{app:proof-hmsc-exis-bounded}.
Basically, one computes the bound for the BMSC of every vertex in $H$ and takes the maximum.
This works since every MSC of $H$ is a concatenation of individual BMSCs which can be scheduled in a way that the channels are empty after each BMSC.

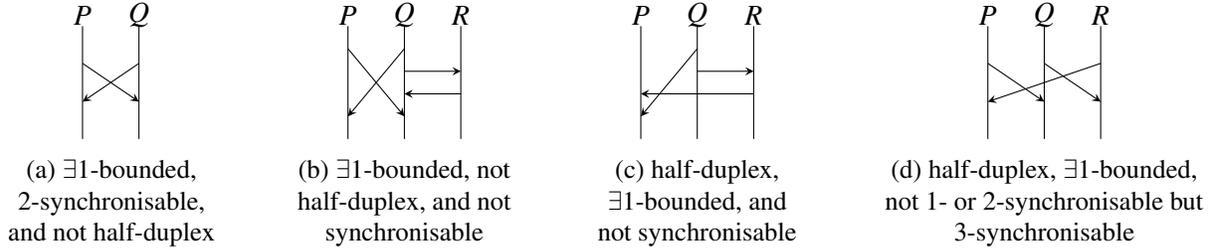
\begin{figure}[t]
\begin{subfigure}[t]{0.18\linewidth}
\centering
    \begin{tikzpicture}[scale=1.5]
        \coordinate (Q1) at (-0.25,0);
\coordinate (Q2) at (-0.25,-1);
\draw(Q2) -- (Q1)node[pos=1.1,scale=1]{$P$};
\coordinate (P1) at (0.25,0);
\coordinate (P2) at (0.25,-1);
\draw(P2) -- (P1)node[pos=1.1,scale=1]{$Q$};
\draw[-stealth] ($(P1)!0.33!(P2)$) -- node[above,scale=1,pos=0.5]{} ($(Q1)!0.67!(Q2)$);
\draw[-stealth] ($(Q1)!0.33!(Q2)$) -- node[above,scale=1,pos=0.5]{} ($(P1)!0.67!(P2)$);
     \end{tikzpicture}
    \caption{$\exists 1$-bounded, $2$-synchronisable, and not half-duplex}
    \label{fig:ex-2-sync-ex-1-bounded-not-half-duplex}
\end{subfigure}
\hfill
\begin{subfigure}[t]{0.18\linewidth}
\centering
    \begin{tikzpicture}[scale=1.5]
    \coordinate (P1) at (-0.5,0);
\coordinate (P2) at (-0.5,-1);
\draw(P2) -- (P1)node[pos=1.1,scale=1]{$P$};
\coordinate (Q1) at (0.0,0);
\coordinate (Q2) at (0.0,-1);
\draw(Q2) -- (Q1)node[pos=1.1,scale=1]{$Q$};
\coordinate (R1) at (0.5,0);
\coordinate (R2) at (0.5,-1);
\draw(R2) -- (R1)node[pos=1.1,scale=1]{$R$};
\draw[-stealth] ($(P1)!0.20!(P2)$) -- node[above,scale=1,pos=0.5]{} ($(Q1)!0.80!(Q2)$);
\draw[-stealth] ($(Q1)!0.40!(Q2)$) -- node[above,scale=1,pos=0.5]{} ($(R1)!0.40!(R2)$);
\draw[-stealth] ($(R1)!0.60!(R2)$) -- node[above,scale=1,pos=0.5]{} ($(Q1)!0.60!(Q2)$);
\draw[-stealth] ($(Q1)!0.20!(Q2)$) -- node[above,scale=1,pos=0.5]{} ($(P1)!0.80!(P2)$);
     \end{tikzpicture}
    \caption{$\exists 1$-bounded, not half-duplex,  and not synchronisable}
    \label{fig:non-sync-bmsc}
\end{subfigure}
\hfill
\begin{subfigure}[t]{0.18\linewidth}
\centering
    \begin{tikzpicture}[scale=1.5]
    \coordinate (P1) at (-0.5,0);
\coordinate (P2) at (-0.5,-1);
\draw(P2) -- (P1)node[pos=1.1,scale=1]{$P$};
\coordinate (Q1) at (0.0,0);
\coordinate (Q2) at (0.0,-1);
\draw(Q2) -- (Q1)node[pos=1.1,scale=1]{$Q$};
\coordinate (R1) at (0.5,0);
\coordinate (R2) at (0.5,-1);
\draw(R2) -- (R1)node[pos=1.1,scale=1]{$R$};
\draw[-stealth] ($(Q1)!0.20!(Q2)$) -- node[above,scale=1,pos=0.5]{$$} ($(P1)!0.80!(P2)$);
\draw[-stealth] ($(Q1)!0.40!(Q2)$) -- node[above,scale=1,pos=0.5]{$$} ($(R1)!0.40!(R2)$);
\draw[-stealth] ($(R1)!0.60!(R2)$) -- node[above,scale=1,pos=0.75]{$$} ($(P1)!0.60!(P2)$);
     \end{tikzpicture}
    \subcaption{half-duplex, $\exists 1$-bounded, and not synchronisable}
    \label{fig:half-duplex-B-bounded-non-sync}
\end{subfigure}
\hfill
\begin{subfigure}[t]{0.27\linewidth}
\centering
    \begin{tikzpicture}[scale=1.5]
    \coordinate (P1) at (-0.5,0);
\coordinate (P2) at (-0.5,-1);
\draw(P2) -- (P1)node[pos=1.1,scale=1]{$P$};
\coordinate (Q1) at (0.0,0);
\coordinate (Q2) at (0.0,-1);
\draw(Q2) -- (Q1)node[pos=1.1,scale=1]{$Q$};
\coordinate (R1) at (0.5,0);
\coordinate (R2) at (0.5,-1);
\draw(R2) -- (R1)node[pos=1.1,scale=1]{$R$};
\draw[-stealth] ($(P1)!0.33!(P2)$) -- node[above,scale=1,pos=0.5]{$$} ($(Q1)!0.67!(Q2)$);
\draw[-stealth] ($(Q1)!0.33!(Q2)$) -- node[above,scale=1,pos=0.5]{$$} ($(R1)!0.67!(R2)$);
\draw[-stealth] ($(R1)!0.33!(R2)$) -- node[above,scale=1,pos=0.5]{$$} ($(P1)!0.67!(P2)$);
     \end{tikzpicture}
    \subcaption{half-duplex, $\exists 1$-bounded, not $1$- or $2$-synchronisable but $3$-synchronisable}
    \label{fig:not-1-sync-k-sync-B-bounded-half-duplex.bmsc.tikz}
\end{subfigure}
\caption{BMSCs which satisfy different channels restrictions}
\label{fig:examples-H}
\end{figure}

\begin{example}\hypouse{\emph{\hypo{H2}}: $\exists B$-bounded, $k$-synchronisable, and not half-duplex.}
\label{ex:ex-2-sync-ex-1-bounded-not-half-duplex}
Consider the BMSC in \cref{fig:ex-2-sync-ex-1-bounded-not-half-duplex}.
It is $\exists 1$-bounded as there is one message per channel, $2$-synchronisable since the message exchange can be split into one phase of two sends and two subsequent receives and not half-duplex because both messages can traverse their channel at the same time.
\end{example}

\begin{example}\hypouse{\emph{\hypo{H3}}: $\exists B$-bounded, not half-duplex, and not synchronisable.}
\label{ex:HMSC-non-sync}
It is obvious that the BMSC~$M$ in \cref{fig:non-sync-bmsc} is not half-duplex.
We show that $M$ is not $k$-synchronous for any~$k$.
Let us denote the event nodes for each process $\procA$ with $p_1,\,\ldots$ as ordered by the total process order.
It is straightforward that one of $p_1$ and $q_1$ has to be part of the first $k$-exchange.
However, since the respective corresponding reception happens after the other's event node, both have to be a part of the first $k$-exchange.
Since these receive event nodes (transitively) depend on all other event nodes, all event nodes have to be part of a single $k$-exchange for $M$.
However, $\procC$ first has to receive from $\procB$ in order to send back to it and therefore, there is no single $k$-exchange for $M$ and $M$ is not $k$-synchronous for any~$k$.
\end{example}

\begin{example}\hypouse{\emph{\hypo{H4}}: half-duplex, $\exists B$-bounded, and not synchronisable.}
\label{ex:half-duplex-B-bounded-non-sync}
Let us consider the BMSC in
\cref{fig:half-duplex-B-bounded-non-sync}.
It is straightforward that it is half-duplex and existentially $1$-bounded.
However, it is not $k$-synchronisable for any~$k$.
In particular, the first and last event node (of any total order induced by the BMSC) must belong to the same message exchange but two more linearly dependent message exchanges need to happen in between.
\end{example}

\begin{example}\hypouse{\emph{\hypo{H5}}: half-duplex, $\exists B$-bounded, $k$-synchronisable but not $1$-synchronisable.}
\label{ex:not-1-sync-k-sync-B-bounded-half-duplex.bmsc.tikz}
Consider the BMSC in
\cref{fig:not-1-sync-k-sync-B-bounded-half-duplex.bmsc.tikz}.
It is easy to see that it is not $1$- or $2$-synchronisable but $3$-synchronisable, half-duplex and existentially $1$-bounded.
Note that it is straightforward to amend the example such that it is still half-duplex but the parameters $B$ and $k$ need to be increased.
\end{example}

\begin{lemma}\label{lm:exis-1-sync-hmsc-half-duplex}
\hypouse{\normalfont $\hyporounded{H6}$:}
Every $1$-synchronisable HMSC is half-duplex.
\end{lemma}
Intuitively, in any BMSC of an HMSC, every send event node has a corresponding receive event node.
Therefore, a message that has been sent needs to have been received directly afterwards and the per-process order is total so any process has to receive a message before it sends a message back.
The full proof can be found in \cref{app:proof-exis-1-sync-half-duplex}.
 \section{Multiparty Session Types}
\label{sec:mst}

In this section, we recall global types from Multiparty Session Types (MSTs) as a way to specify protocols.
We present an embedding for MSTs into HMSCs, prove it correct, and use it to show that MSTs are half-duplex, existentially $1$-bounded, and $1$-synchronisable.

\subsection{Specifying Protocols with Global Types}

We now define global types in the framework of MSTs as a syntax for protocol specifications.
The syntax of global types is defined following classical MST frameworks \cite[Def.~3.2]{DBLP:journals/pacmpl/ScalasY19}.
The calculus focuses on the core message-passing primitives of asynchronous MSTs and does not incorporate features like subsessions or delegation.
However, it does incorporate a recent generalisation that allow a sender to send to different receivers upon branching \cite{DBLP:conf/concur/MajumdarMSZ21}.

\begin{definition}[Syntax of Global Types \cite{DBLP:conf/concur/MajumdarMSZ21}]
\emph{Global types for MSTs} are defined by the grammar:
    \begin{grammar}
     G \is
       0
     | \sum_{i ∈ I} \msgFromTo{\procA}{\procB_i}{\val_i.G_i}
     | μ t. G
     | t
    \end{grammar}
\end{definition}

An expression $\msgFromTo{\procA}{\procB}{\val}$ stands for a send and receive event: $\snd{\procA}{\procB}{\val}$ and $\rcv{\procA}{\procB}{\val}$.
Since global types always specify send and the receive events together, they specify complete \channelcompliant sequences of events.
For a \emph{choice}, the sender process decides which branch to take and
each branch of a choice needs to be uniquely distinguishable ($∀ i,j ∈ I.\, i≠j ⇒ \procB_i \neq \procB_j \lor \val_i ≠ \val_j$).
If there is a single alternative (and no actual choice), we omit writing the sum operator.
Loops are encoded by the least fixed point operator and recursion must be guarded, i.e., there is at least one message between $μt$ and~$t$.
We assume, without loss of generality, that all occurrences of recursion variables $t$ are bound and every variable $t$ is distinct.
Recursion only happens at the tail (and there is no additional parameter) and, therefore, the language of a global type can be defined with an automaton -- as expected by following the structure of a global type, splitting the message exchanges into send and receive events while not only accounting for finite but also infinite executions.
We give one language as example and refer to \cref{app:semantics-mst} for a formal~definition.

\begin{example}
\label{ex:hmsc-less-order}
The type language for the global type in \cref{sec:intro},
$
μt. \;
(
   \msgFromTo{\procA}{\procB}{\mathit{cons}}.\,t
   \, + \,
   \msgFromTo{\procA}{\procB}{\mathit{nil}}.\,\msgFromTo{\procB}{\procA}{\mathit{ack}}.\,0
)
$,
is the union of a set of finite executions and infinite executions:
\[
\bigl(
        \snd{\procA}{\procB}{\mathit{cons}}.\,
        \rcv{\procA}{\procB}{\mathit{cons}}
    \bigr)^*
    . \;
    \snd{\procA}{\procB}{\mathit{nil}}.\,
    \rcv{\procA}{\procB}{\mathit{nil}}.\,
    \snd{\procB}{\procA}{\mathit{ack}}.\,
    \rcv{\procB}{\procA}{\mathit{ack}}
\quad \text{ and } \quad
\bigl(
        \snd{\procA}{\procB}{\mathit{cons}}.\,
        \rcv{\procA}{\procB}{\mathit{cons}}
    \bigr)^\omega
\]
\end{example}

\begin{remark}
For readers familiar with MSTs, it may be strange that we do not define local types.
In fact, one correctness criterion for local types requires that their composition generates the same language as the original global type.
All channel restrictions are defined using languages, so it suffices to consider global types for our purposes.
\end{remark}

\begin{example}
\label{ex:reordering-necessary}
Consider the global type: $\msgFromTo{\procA}{\procB}{\val₁}.\msgFromTo{\procC}{\procD}{\val₂}$.
The type language for this type contains only the word $\snd{\procA}{\procB}{\val₁}. \rcv{\procA}{\procB}{\val₁}. \snd{\procC}{\procD}{\val₂}. \rcv{\procC}{\procD}{\val₂}$.
On the other hand, if we want to describe the same protocol with a HMSC,
it always allows any permutation of
the events where
$\snd{\procA}{\procB}{\val₁}$ occurs before
$\rcv{\procA}{\procB}{\val₁}$ and $\snd{\procC}{\procD}{\val₂}$ before
$\rcv{\procC}{\procD}{\val₂}$.
\end{example}

Intuitively, some events in a distributed setting shall not be ordered since they are independent, e.g., happen on different processes as in the previous example.
To this end, we recall an indistinguishability relation $\interswap$ that captures the reordering allowed by CSMs with FIFO channels (which will be defined in \cref{sec:csm}).
In MSTs, similar reordering rules are applied (e.g.,
\cite[Def.~3.2 and 5.3]{DBLP:journals/jacm/HondaYC16}).

\begin{definition}[Indistinguishability relation $\interswap$ \cite{DBLP:conf/concur/MajumdarMSZ21}]
Let ${\interswap_i} \subseteq \Sigma^* \times \Sigma^*$, for $i\geq 0$, be a family of indistinguishability relations.
For all $w\in\Sigma^*$, we have $w \interswap_0 w$.
For $i=1$, we define:
{ \small
\begin{enumerate}[label=(\arabic*)]
\item
If $\procA ≠ \procC$, then $ w.\snd{\procA}{\procB}{\val}.\snd{\procC}{\procD}{\val'}.u
 \; \interswap_{1} \;
 w.\snd{\procC}{\procD}{\val'}.\snd{\procA}{\procB}{\val}.u.
$ \item
If $\procB ≠ \procD$, then $ w.\rcv{\procA}{\procB}{\val}.\rcv{\procC}{\procD}{\val'}.u
 \; \interswap_{1} \;
 w.\rcv{\procC}{\procD}{\val'}.\rcv{\procA}{\procB}{\val}.u.
$ \item
If $\procA ≠ \procD \land (\procA ≠ \procC ∨ \procB ≠ \procD)
$, 
then $ w.\snd{\procA}{\procB}{\val}.\rcv{\procC}{\procD}{\val'}.u
 \; \interswap_{1} \;
 w.\rcv{\procC}{\procD}{\val'}.\snd{\procA}{\procB}{\val}.u.
$ \item
If $\card{w \wproj_{\snd{\procA}{\procB}{\_}}} >
    \card{w \wproj_{\rcv{\procA}{\procB}{\_}}}$,
then $ w.\snd{\procA}{\procB}{\val}.\rcv{\procA}{\procB}{\val'}.u
 \; \interswap_{1} \;
 w.\rcv{\procA}{\procB}{\val'}.\snd{\procA}{\procB}{\val}.u.
$ \end{enumerate}
}
Let $w, w', w''$ be sequences of events such that $w \interswap_1 w'$ and $w' \interswap_i w''$ for some~$i$.
Then, $w \interswap_{i+1} w''$.
We define $w \interswap u$ if there is $n$ such that $w \interswap_n u$.
\end{definition}

This formalises how  messages can be swapped for finite executions of protocols.
The infinite case requires special technical treatment for which we refer to the work by Majumdar et al.~\cite{DBLP:conf/concur/MajumdarMSZ21}.

The relation is lifted to languages as expected.
For a language $L$, we have:

\smallskip
{ \small
\hspace{.15\textwidth}
$
  \interswaplang(L) = \left\{ w' \mid \bigvee
    \begin{array}{l}
    w' \in \Alphabet^* \land ∃ w ∈ \Alphabet^*. \; w \in L \text{ and } w' \interswap w \\
    w' ∈ \Alphabet^ω \land \exists w \in \Alphabet^\omega. \; w \in 
    L \text{ and } w' \preceq_\interswap^\omega w
  \end{array} \right\}.
$
}
\smallskip

The indistinguishability relation $\interswap$ does not change the order of send and receive events of a single process.
The relation $\interswap$ captures all reorderings which naturally appear when global types from MSTs are implemented with CSMs.
For a global type $G$, its semantics is given by its \emph{execution language} $\interswaplang(\lang(G))$.
Furthermore, the indistinguishability relation captures exactly the events that are independent in any HMSC.
Phrased differently, HMSC include these reorderings by design.

\begin{lemma}
\label{lm:hmsc-closed-indistinguishability-relation}
Let $H$ be any HMSC.
Then, $\lang(H) = \interswaplang(\lang(H))$.
\end{lemma}
The proof is in \cref{app:hmsc-indistinguishability-relation}.
A similar result for CSMs has previously been proven \cite[Lemma~21]{DBLP:conf/concur/MajumdarMSZ21}.

\begin{theorem}
\label{lm:protocols-closed-indistinguishability-relation}
For \channelcompliant words,
the indistinguishability relation
$\interswap$ preserves satisfaction of half-duplex communication, existential $B$-boundedness, and $k$-synchronisability.
\end{theorem}
The proof can be found in \cref{app:protocols-closed-indistinguishability-relation}.

\subsection{Encoding Global Types from MSTs into HMSCs}
\label{sec:mst-to-hmsc}

Global types from MSTs can be turned into HMSCs while preserving the protocol they specify.
In this step, we account for the orders than can and cannot be enforced in an asynchronous point-to-point setting with the indistinguishability relation $\interswap$.
The main difference between the automata-based semantics of global types from MSTs
and the semantics of HMSCs is that
    an automaton carries the events on the edges and
    an HMSC carries events as labels of the event nodes in the BMSCs associated with the vertices.

In the translation, we use the following notation.
$M_\emptyset$ is the empty BMSC ($\eventnodes = ∅$) and
$M( \msgFromTo{\procA}{\procB}{\val} )$ is the BMSC with two event nodes: $e₁$, $e₂$ such that
    $f(e₁) = e₂$,
    $l(e₁) = \snd{\procA}{\procB}{\val}$, and
    $l(e₂) = \rcv{\procA}{\procB}{\val} \,$.

From a global type $G$, we construct an HMSC $H(G) = (V,\edges,v^I,V^T,μ,λ)$~with

\smallskip
\begin{small}
$
\begin{array}{llll}
V     = \; & \set{G' \; \mid \; G' \text{ is a subterm of } G } \; ∪ \; \set{ (\sum_{i ∈ I} \msgFromTo{\procA}{\procB_i}{\val_i}.G_i, j) \; \mid \;
\sum_{i ∈ I} \msgFromTo{\procA}{\procB_i}{\val_i}.G_i  \text{ occurs in } G ∧ j∈ I }
        \phantom{some }
    \vspace{1.5ex}
        \\
\edges = \; & \set{ (μt.G', G') \; \mid \; μt.G'  \text{ occurs in } G }  \; ∪ \;  \set{ (t, μt.G') \; \mid \; t, μt.G'  \text{ occurs in } G }
        \vspace{0.5ex}
            \\
           &  ∪ \; \set{ (\sum_{i ∈ I} \msgFromTo{\procA}{\procB_i}{\val_i}.G_i, (\sum_{i ∈ I} \msgFromTo{\procA}{\procB_i}{\val_i}.G_i,j))  \; \mid \;
(\sum_{i ∈ I} \msgFromTo{\procA}{\procB_i}{\val_i}.G_i,j) ∈ V}
        \vspace{0.5ex}
          \\
          & ∪ \; \set{ ( (\sum_{i ∈ I} \msgFromTo{\procA}{\procB_i}{\val_i}.G_i, j), G_j) \; \mid \;
(\sum_{i ∈ I} \msgFromTo{\procA}{\procB_i}{\val_i}.G_i, j) ∈ V }
    \vspace{1.5ex}
          \\
v^I  = \; & G \qquad \;
V^T  = \;  \set{0} \qquad \;
μ(v) = \;
\begin{cases}
    M( \msgFromTo{\procA}{\procB_i}{\val_j})   & \text{if } v = (\sum_{i ∈ I} \msgFromTo{\procA}{\procB_i}{\val_i}.G_i\}, j) \\
    M_\emptyset                     & \text{otherwise}
\end{cases}
\end{array}
$
\end{small}
\smallskip

This translation does not yield the HMSC with the least number of vertices since vertices with a single successor could be merged to form larger BMSCs.
Here, every BMSC contains at most one message exchange.
We obtain the following correctness statement for the~embedding:

\begin{theorem}
\label{thm:correctnessMPSTtoHMSC}
For any global type $G$,
it holds that
$\lang(G) ⊆ \lang(H(G))$ and
$\interswaplang(\lang(G)) = \interswaplang(\lang(H(G)))$.
\end{theorem}

We provide the technical developments to show
\cref{thm:correctnessMPSTtoHMSC}
in \cref{app:relation-mst-hmsc}.

\begin{remark}
The first part of \cref{thm:correctnessMPSTtoHMSC} uses $⊆$ instead
of $=$ as HMSCs do not order indistinguishable events and we consider the type language of $G$.
\cref{ex:reordering-necessary} shows that using the execution language rather than the type language in the second part is inevitable for equality and does not weaken the claim.
\end{remark}

\subsection{Channel Restrictions of Global Types}
\label{sec:channeluse-mst}

\begin{example}\hypouse{\emph{\hypo{H7}}: half-duplex, $\exists 1$-bounded, $1$-synchronisable but not in MSTs.}
\label{ex:non-MST-1-bounded-1-sync-half-duplex}

\noindent
\begin{minipage}{\textwidth}
\begin{wrapfigure}{r}{0.24\textwidth}
\vspace{-7ex}
  \begin{center}
    \includegraphics[width=0.21\textwidth]{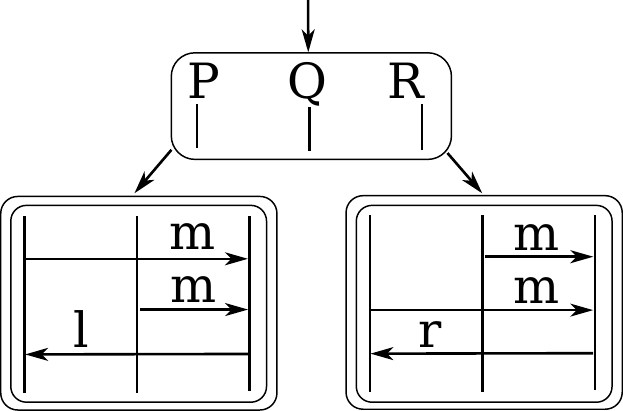}
  \end{center}
\vspace{-2.5ex}
  \caption{half-duplex, $\exists 1$-bounded, and $1$-synchronisable but not expressible in MSTs}
  \label{fig:non-MST-1-bounded-1-sync-half-duplex}
\end{wrapfigure}\vspace{1ex}
Consider the HMSC in \cref{fig:non-MST-1-bounded-1-sync-half-duplex}.
It is straightforward that it is half-duplex, existentially $1$-bounded, and $1$-synchronisable.
Both $\procA$ and $\procB$ send the same message to $\procC$ independently in each branch.
Intuitively, $\procC$ chooses which branch to take by the order it decides to receive both messages.
Subsequently, it notifies $\procB$ about this choice.
Such a communication pattern cannot be expressed in the MST framework.
If one tried to model it with $\procC$ actually choosing the branch, $l$ and $r$ would always occur before the receptions so the languages are different.
\end{minipage}
\end{example}

We show that protocols specified as global types satisfy all discussed channel restrictions (with the minimal reasonable parameter).

\begin{theorem}\hypouse{\emph{$\hyporounded{H8}$:}}
\label{cor:MST-ex-1-sync-and-ex-1-bounded}
The execution language $\interswaplang(\lang(G))$ is half-duplex, existentially $1$-bounded, and $1$-synchronisable for any global type $G$.
\end{theorem}

The proof uses the embedding to obtain an HMSC built of BMSCs with at most one message exchange and exploits previously shown properties about HMSCs. Details can be found in \cref{app:mst-channel-use}.

\begin{remark}[Choreography automata are half-duplex, $\exists 1$-bounded, and $1$-synchronisable]
\phantom{}\\
In this section, we looked at MSTs which are rooted in process algebra.
With \emph{choreography automata} \cite{DBLP:conf/coordination/BarbaneraLT20}, a~similar concept has been studied from automata theory perspective.
Basically, a protocol specification is an automaton whose transitions are labelled by $\msgFromTo{\procA}{\procB}{\val}$.
In contrast to global types from MSTs, they do not impose constraints on choice, i.e., there does not need to be a unique process chooses which branch to take next and do not employ an indistinguishability relation but require to explicitly spell out all possible reorderings.
This feature can lead to complications w.r.t.\ implementing such protocols but does not change the satisfaction of channel restrictions.
In fact, protocols specified by choreography automata are also half-duplex, existentially $1$-bounded, and $1$-synchronisable.
\end{remark}
 \section{Communicating State Machines}
\label{sec:implementing-protocols}

In this section, we first present communicating state machines (CSMs) as formal model for distributed processes which communicate messages asynchronously via reliable point-to-point FIFO channels.
If a~CSM implements a protocol specification, both languages are the same -- modulo $\interswap$ which does not alter satisfaction of channel restrictions.
This entails that CSMs implementing protocol specifications satisfy the same channel restrictions as presented in previous sections.
Here, we investigate the channel restrictions of general CSMs which might not implement a protocol specified as global type or HMSC.

\begin{definition}[Communicating state machines]
\label{sec:csm}
A \emph{state machine} $A = (Q, \Delta, \delta, q_{0}, F)$ is a $5$-tuple where
$Q$ is a finite set of states,
$\Delta$ is an alphabet,
$\delta \subseteq Q \times (\Sigma \union \set{\emptystring}) \times Q$ is a transition relation,
$q_{0}\in Q$ is an initial state, and
$F \subseteq Q$ is a set of final states.
We write $q \xrightarrow{a} q'$ for $(q, a, q')\in\delta$.
A \emph{run} of $A$ is a sequence $\rho = q_0\xrightarrow{w_0} q_1 \xrightarrow{w_1} \ldots$,
with $q_i~\in~Q$ and $w_i\in \Delta \union \set{\emptystring}$ for $i\geq 0$,
such that $q_0$ is the initial state, and for each $i\geq 0$, it holds that $(q_i, w_i, q_{i+1})\in\delta$.
The \emph{trace} of the run
is the finite or infinite word $w_0w_1\ldots\in \Sigma^\infty$.
The \emph{path} of the run is the finite or infinite sequence $q_0q_1 \ldots\in Q^\infty$.
A run is called maximal if it is infinite or ends at a final state.
Accordingly, the corresponding trace and path are called maximal.
The \emph{language} $\lang(A)$ of $A$ is the set of its maximal~traces.

We call $\mathcal{A} = \CSM{A}$ a \emph{communicating state machine} (CSM)  over $\Procs$ and~$\MsgVals$ if
${A}_\procA$
is a finite state machine
with alphabet~$\Sigma_\procA$ for every $\procA\in\Procs$.
The state machine for $\procA$ is denoted by $(Q_\procA, \Sigma_\procA, \delta_\procA, q_{0, \procA}, F_\procA)$.
Intuitively, a CSM allows a set of state machines, one for each process in $\Procs$,
to communicate by sending and receiving messages.
For this, each pair of processes $\procA, \procB\in \Procs$, $\procA \neq\procB$, is connected by two directed \emph{message channels}.
A transition $q_{\procA} \xrightarrow{\snd{\procA}{\procB}{\val}} q'_{\procA}$ in the state machine of $\procA$ denotes that $\procA$ sends message $\val$ to $\procB$ if $\procA$ is in the state~$q$ and changes its local state to~$q'$.
The channel $\channel{\procA}{\procB}$ is appended by message~$\val$.
For receptions, a transition $q_{\procB} \xrightarrow{\rcv{\procA}{\procB}{\val}} q'_{\procB}$ in the state machine of $\procB$
corresponds to $\procB$ retrieving the message $\val$ from the head of the channel when its local state is $\hat{q}$ which is updated
to $\hat{q}'$.
The run of a CSM always starts with empty channels and each finite state machine is its respective initial state.
The formalisation of this intuition is standard and can be found in \cref{app:semantics-csm}.
\end{definition}

As for HMSCs, the language of a CSM is closed under~$\interswap$.
\begin{lemma}
[\cite{DBLP:conf/concur/MajumdarMSZ21}, Lemma~21]
\label{lm:csm-closed-under-interswap}
Let $𝓐$ be a CSM.
Then $\lang(𝓐) = \interswaplang(\lang(𝓐))$.
\end{lemma}

\smallskip\noindent\textbf{Implementing Protocol Specifications.}
Given a protocol specification, one can try to generate an implementation which admits the same language.
This problem is known as implementability or realisability in the HMSC setting \cite{DBLP:journals/jcss/GenestMSZ06,DBLP:journals/tcs/AlurEY05,DBLP:journals/tcs/Lohrey03}.
In the MST setting \cite{DBLP:conf/popl/HondaYC08}, this is done in two steps.
First, the global type is projected on to local types.
Second, a type system ensures that the implementations follow the local types, i.e., a refinement~check.
However, one can design CSMs from scratch that yield systems which cannot be captured by protocol specifications like HMSCs or global types from MSTs.

\subsection{Channel Restrictions of CSMs}
\label{sec:understanding-channel-use}

We say that an CSM is half-duplex, existentially $B$-bounded, or $k$-synchronisable respectively if its language~is.
Again, we follow the outline presented in \cref{fig:overview-csm}.

\begin{example}
\hypouse{\emph{\hypo{C1}:}} half-duplex, $\exists B$-bounded, and $k$-synchronisable.
The CSM in \cref{fig:intro-csm} is $\exists 1$-bounded, $1$-synchronisable, and half-duplex.
\end{example}

Any BMSC can easily be implemented with an CSM by simple letting each process follow its linear trajectory of eventnodes.
We call this projection.
Therefore, we can use three of the BMSCs presented in \cref{fig:examples-H} to show the hypotheses for CSMs:

\begin{example}
For, \hypouse{\emph{\hypo{C2}}}, the projection of \cref{fig:half-duplex-B-bounded-non-sync} (used to show \emph{\hypo{H4}}) is
half-duplex, $\exists B$-bounded, and not synchronisable.
For \hypouse{\emph{\hypo{C3}}}, the projection of \cref{fig:non-sync-bmsc} (used to show \emph{\hypo{H3}})
is $\exists B$-bounded, not half-duplex, and not synchronisable.
For \hypouse{\emph{\hypo{C4}}}, the projection of \cref{fig:ex-2-sync-ex-1-bounded-not-half-duplex} (used to show \emph{\hypo{H2}})
is $\exists B$-bounded, $k$-synchronisable, and not half-duplex.
\end{example}

\begin{example}\label{ex:ex-1-sync-but-non-ex-bounded}
\hypouse{
\emph{\hypo{C5}:}
$k$-synchronisable, not half-duplex and not $\exists$-bounded; \\
\emph{\hypo{C6}:}
$k$-synchronisable, half-duplex and not $\exists$-bounded.
}

\noindent
\begin{minipage}{\textwidth}
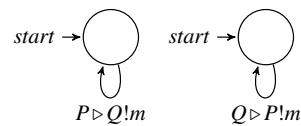
\begin{wrapfigure}[10]{r}{0.27\textwidth}
\vspace{-4.5ex}
\centering
\resizebox{0.26\textwidth}{!}{
\begin{tikzpicture}[->,>=stealth',shorten >=1pt,auto,node distance=2.7cm,semithick,scale=0.7]
  \tikzstyle{every state}=[circle,draw,text=white]

  \node[initial,state] (A)              {$q_a$};
  \path (A) edge [loop below] node {$\snd{\procA}{\procB}{\val}$} (A);

  \node[initial,state, right of=A] (B)              {$q_b$};
  \path (B) edge [loop below] node {$\snd{\procB}{\procA}{\val}$} (B);
\end{tikzpicture}
 }
\caption{CSM with FSMs for $\procA$ (left) and for $\procB$ (right)} \label{fig:ex-1-sync-but-non-ex-bounded}
\end{wrapfigure}

\smallskip
We consider two CSMs constructed from the state machines in
\cref{fig:ex-1-sync-but-non-ex-bounded}.
For~\emph{\hypo{C5}}, we consider the CSM consisting of both state machines.
It is $1$-synchronisable but not existentially bounded and not half-duplex.
It is $1$-synchronisable because every linearisation can be split into single send events that constitute $1$-exchanges.
It is neither existentially $B$-bounded for any $B$ nor half-duplex since none of the messages will be received so both channels can grow arbitrarily.
For~\emph{\hypo{C6}}, it can easily be turned into a half-duplex CSM by removing one of the send events.
Then, the CSM is $1$-synchronisable and half-duplex but not existentially bounded.
\end{minipage}
\end{example}

This example disproves a result from the literature \cite[Thm.~7.1]{DBLP:conf/cav/LangeY19}, which states that every $k$-synchro\-nisable system is existentially $B$-bounded for some $B$ and has been cited recently as part of a summary~\cite[Prop.~41]{DBLP:conf/concur/BolligGFLLS21}.
In the proof, it is neglected that unreceived messages remain in the channels after a message exchange.
Our example satisfies their assumption that CSMs do not have states with mixed choice, i.e., each state either is final, has send options to choose from, or receive options to choose from.
We do not impose any assumptions on mixed choice in this work.
Still, all the presented examples do not have states with mixed choice so the presented relationships also hold for this subset of~CSMs.
\begin{corollary}
 Existential $B$-boundedness and $k$-synchronisability for CSMs are incomparable.
\end{corollary}
The previous result follows immediately from the CSMs constructed in \cref{ex:ex-1-sync-but-non-ex-bounded}.
Our result considers the point-to-point FIFO setting. For the mailbox setting, the analogous question is an open problem~\cite{DBLP:conf/cav/BouajjaniEJQ18}.

\smallskip\noindent\textbf{Turing-powerful Encodings.}
On the one hand, it is well-known that CSMs are Turing-complete \cite{DBLP:journals/jacm/BrandZ83} and C{\'{e}}c{\'{e}} and Finkel \cite[Thm.~36]{DBLP:journals/iandc/CeceF05} showed that half-duplex communication does not impair expressiveness of CSMs with more than two processes.
On the other hand, each of existential $B$-boundedness and $k$-synchronisability render some verification questions decidable.
Therefore, the encodings of Turing-completeness \cite{DBLP:journals/jacm/BrandZ83,DBLP:journals/iandc/CeceF05} are examples for CSMs which are not existentially $B$-bounded for any $B$ nor $k$-synchronisable for any~$k$ and either half-duplex ($\hypo{C7}$) or not half-duplex ($\hypo{C8}$).
 \section{Related Work}
\label{sec:related}

We now cover related work which is not already cited in the earlier sections.

The origins of MSTs date back to 1993 when Honda et al.~\cite{DBLP:conf/concur/Honda93} proposed a binary version for typing communication in the domain of process algebra.
In 2008, Honda et al.~\cite{DBLP:conf/popl/HondaYC08} generalised the idea to multiparty systems.
While the connection of MSTs and CSMs has been studied soon after MSTs had been proposed \cite{DBLP:journals/corr/abs-1203-0780,DBLP:conf/esop/DenielouY12},
we provide, to the best of our knowledge, the first formal connection of MSTs to HMSCs, even though HMSC-like visualisations have been used in the community of session types, e.g.~\cite[Fig.~1]{Carbone2005ATB}, \cite[Figs.~1 and 2]{DBLP:journals/jacm/HondaYC16}.
For binary session types, it is known how to compute the bound~$B$ of universally $B$-bounded types \cite{DBLP:conf/eurosys/FahndrichAHHHLL06,DBLP:journals/jfp/GayV10}.
Lange et al.~\cite{DBLP:conf/cav/LangeY19} proposed $k$-multiparty consistency ($k$-MC) for CSMs as extension of multiparty consistency for MSTs.
We did not consider $k$-MC in this work for two reasons.
First, they assume an existential bound (of $k$) on channels.
Second, as an extension of multiparty consistency, $k$-MC focuses on implementability rather than channel restrictions.

HMSCs and variants thereof have been extensively studied~\cite{DBLP:conf/ac/GenestMP03,DBLP:conf/acsd/GenestM05,DBLP:conf/concur/GazagnaireGHTY07,DBLP:journals/tosem/RoychoudhuryGS12}.The connection to CSMs has been investigated in particular for different forms of implementability \cite{DBLP:journals/jcss/GenestMSZ06,DBLP:journals/tcs/AlurEY05,DBLP:journals/tcs/Lohrey03}, also called realisability \cite{DBLP:journals/tcs/AlurEY05}, which is undecidable in general \cite{DBLP:conf/ac/GenestMP03,DBLP:journals/tcs/AlurEY05}, and implied scenarios  \cite{DBLP:conf/fase/Muccini03,DBLP:conf/fase/MooijGR05} which arise when implementing HMSCs with CSMs.
Several restrictions to check implementability adopted a limited form of choice \cite{DBLP:conf/tacas/Ben-AbdallahL97,NO-DBLP-wrong-local-choice,DBLP:conf/fase/Muccini03,DBLP:conf/fase/MooijGR05,DBLP:conf/sefm/DanHC10}
which is similar to the one in global types from MSTs.
For more details, we refer to work by Majumdar et al.~\cite{DBLP:conf/concur/MajumdarMSZ21}.

While we consider finite state machines as model for processes, research has also been conducted on communicating systems where processes are given more computational power, e.g., pushdown automata \cite{DBLP:journals/corr/abs-1209-0359,DBLP:journals/tcs/TouiliA10,DBLP:conf/atva/AiswaryaGK14}.
However, as noted before, our setting is already Turing-powerful.
In \cref{sec:alg-verification}, we surveyed how channel restrictions can yield decidability.
Incomplete approaches consider subclasses which enable the effective computation of symbolic representations (of channel contents) for reachable states~\cite{
DBLP:conf/sas/BoigelotGWW97,DBLP:journals/mst/Kocher21}.
Other approaches change the semantics of channels, e.g., by making them lossy
\cite{
DBLP:conf/cav/AbdullaBJ98,
DBLP:conf/lics/AbdullaAA16,DBLP:journals/mst/Kocher21},
input-bounded \cite{DBLP:conf/concur/BolligFS20},
or by restricting the communication topology \cite{DBLP:journals/acta/PengP92,DBLP:conf/tacas/TorreMP08}.

While we build on the most recent definitions of synchronisability~\cite{DBLP:conf/fossacs/GiustoLL20}, we refer to the work by Finkel and Lozes~\cite{DBLP:conf/icalp/FinkelL17} and Bouajjani et al.~\cite{DBLP:conf/cav/BouajjaniEJQ18} for earlier work on synchronisability.
Bollig et.\ al \cite{DBLP:conf/concur/BolligGFLLS21} studied the connection of different notions of synchronisability for MSCs and MSO logic which yields interesting decidability results.
We refer to their work for more details but briefly point to the slightly different use of terminology:
$k$-synchronisability is called weak ($k$-)synchronisability by Bollig where the omission of $k$ indicates a system is synchronisable for some $k$;
while strong ($k$-)synchronisability does solely apply to the mailbox setting.
 \section{Conclusion}
\label{sec:conclusion}

We presented a comprehensive comparison of half-duplex, existential $B$-bounded, and \mbox{$k$-syn}\-chro\-nis\-able communication.
We showed that the three restrictions are different for CSMs.
For \mbox{HMSCs}, the half-duplex restriction and $k$-syn\-chro\-nis\-ability are different and included in existential \mbox{$B$-boundedness}.
Furthermore, all $1$-synchronisable HMSC-definable languages are half-duplex.
This subclass contains global types from MSTs which are also existentially $1$-bounded.
We established the first formal \mbox{embedding} of~global types from MSTs into HMSCs which can be used to combine insights from both domains for further advances on implementing protocol specifications.
 
\pagebreak

\bibliographystyle{eptcs}

\begin{thebibliography}{10}
\providecommand{\bibitemdeclare}[2]{}
\providecommand{\surnamestart}{}
\providecommand{\surnameend}{}
\providecommand{\urlprefix}{Available at }
\providecommand{\url}[1]{\texttt{#1}}
\providecommand{\href}[2]{\texttt{#2}}
\providecommand{\urlalt}[2]{\href{#1}{#2}}
\providecommand{\doi}[1]{doi:\urlalt{https://doi.org/#1}{#1}}
\providecommand{\eprint}[1]{arXiv:\urlalt{https://arxiv.org/abs/#1}{#1}}
\providecommand{\bibinfo}[2]{#2}

\bibitemdeclare{inproceedings}{DBLP:conf/lics/AbdullaAA16}
\bibitem{DBLP:conf/lics/AbdullaAA16}
\bibinfo{author}{Parosh~Aziz \surnamestart Abdulla\surnameend},
  \bibinfo{author}{C.~\surnamestart Aiswarya\surnameend} \&
  \bibinfo{author}{Mohamed~Faouzi \surnamestart Atig\surnameend}
  (\bibinfo{year}{2016}): \emph{\bibinfo{title}{Data Communicating Processes
  with Unreliable Channels}}.
\newblock In \bibinfo{editor}{Martin \surnamestart Grohe\surnameend},
  \bibinfo{editor}{Eric \surnamestart Koskinen\surnameend} \&
  \bibinfo{editor}{Natarajan \surnamestart Shankar\surnameend}, editors:
  {\slshape \bibinfo{booktitle}{Proceedings of the 31st Annual {ACM/IEEE}
  Symposium on Logic in Computer Science, {LICS} '16, New York, NY, USA, July
  5-8, 2016}}, \bibinfo{publisher}{{ACM}}, pp. \bibinfo{pages}{166--175},
  \doi{10.1145/2933575.2934535}.
\newblock \urlprefix\url{https://doi.org/10.1145/2933575.2934535}.

\bibitemdeclare{inproceedings}{DBLP:conf/cav/AbdullaBJ98}
\bibitem{DBLP:conf/cav/AbdullaBJ98}
\bibinfo{author}{Parosh~Aziz \surnamestart Abdulla\surnameend},
  \bibinfo{author}{Ahmed \surnamestart Bouajjani\surnameend} \&
  \bibinfo{author}{Bengt \surnamestart Jonsson\surnameend}:
  \emph{\bibinfo{title}{On-the-Fly Analysis of Systems with Unbounded, Lossy
  {FIFO} Channels}}.
\newblock In \bibinfo{editor}{Alan~J. \surnamestart Hu\surnameend} \&
  \bibinfo{editor}{Moshe~Y. \surnamestart Vardi\surnameend}, editors: {\slshape
  \bibinfo{booktitle}{Computer Aided Verification, 10th International
  Conference, {CAV}'98, Vancouver, BC, Canada, June 28 - July 2, 1998,
  Proceedings}}.

\bibitemdeclare{inproceedings}{DBLP:conf/atva/AiswaryaGK14}
\bibitem{DBLP:conf/atva/AiswaryaGK14}
\bibinfo{author}{C.~\surnamestart Aiswarya\surnameend}, \bibinfo{author}{Paul
  \surnamestart Gastin\surnameend} \& \bibinfo{author}{K.~Narayan \surnamestart
  Kumar\surnameend} (\bibinfo{year}{2014}): \emph{\bibinfo{title}{Verifying
  Communicating Multi-pushdown Systems via Split-Width}}.
\newblock In \bibinfo{editor}{Franck \surnamestart Cassez\surnameend} \&
  \bibinfo{editor}{Jean{-}Fran{\c{c}}ois \surnamestart Raskin\surnameend},
  editors: {\slshape \bibinfo{booktitle}{Automated Technology for Verification
  and Analysis - 12th International Symposium, {ATVA} 2014, Sydney, NSW,
  Australia, November 3-7, 2014, Proceedings}}, {\slshape
  \bibinfo{series}{Lecture Notes in Computer Science}} \bibinfo{volume}{8837},
  \bibinfo{publisher}{Springer}, pp. \bibinfo{pages}{1--17},
  \doi{10.1007/978-3-319-11936-6\_1}.
\newblock \urlprefix\url{https://doi.org/10.1007/978-3-319-11936-6\_1}.

\bibitemdeclare{article}{DBLP:journals/tse/AlurEY03}
\bibitem{DBLP:journals/tse/AlurEY03}
\bibinfo{author}{Rajeev \surnamestart Alur\surnameend}, \bibinfo{author}{Kousha
  \surnamestart Etessami\surnameend} \& \bibinfo{author}{Mihalis \surnamestart
  Yannakakis\surnameend} (\bibinfo{year}{2003}):
  \emph{\bibinfo{title}{Inference of Message Sequence Charts}}.
\newblock {\slshape \bibinfo{journal}{{IEEE} Trans. Software Eng.}}
  \bibinfo{volume}{29}(\bibinfo{number}{7}), pp. \bibinfo{pages}{623--633},
  \doi{10.1109/TSE.2003.1214326}.
\newblock \urlprefix\url{https://doi.org/10.1109/TSE.2003.1214326}.

\bibitemdeclare{article}{DBLP:journals/tcs/AlurEY05}
\bibitem{DBLP:journals/tcs/AlurEY05}
\bibinfo{author}{Rajeev \surnamestart Alur\surnameend}, \bibinfo{author}{Kousha
  \surnamestart Etessami\surnameend} \& \bibinfo{author}{Mihalis \surnamestart
  Yannakakis\surnameend} (\bibinfo{year}{2005}):
  \emph{\bibinfo{title}{Realizability and verification of {MSC} graphs}}.
\newblock {\slshape \bibinfo{journal}{Theor. Comput. Sci.}}
  \bibinfo{volume}{331}(\bibinfo{number}{1}), pp. \bibinfo{pages}{97--114},
  \doi{10.1016/j.tcs.2004.09.034}.
\newblock \urlprefix\url{https://doi.org/10.1016/j.tcs.2004.09.034}.

\bibitemdeclare{inproceedings}{DBLP:conf/concur/AlurY99}
\bibitem{DBLP:conf/concur/AlurY99}
\bibinfo{author}{Rajeev \surnamestart Alur\surnameend} \&
  \bibinfo{author}{Mihalis \surnamestart Yannakakis\surnameend}
  (\bibinfo{year}{1999}): \emph{\bibinfo{title}{Model Checking of Message
  Sequence Charts}}.
\newblock In \bibinfo{editor}{Jos C.~M. \surnamestart Baeten\surnameend} \&
  \bibinfo{editor}{Sjouke \surnamestart Mauw\surnameend}, editors: {\slshape
  \bibinfo{booktitle}{{CONCUR} '99: Concurrency Theory, 10th International
  Conference, Eindhoven, The Netherlands, August 24-27, 1999, Proceedings}},
  {\slshape \bibinfo{series}{Lecture Notes in Computer Science}}
  \bibinfo{volume}{1664}, \bibinfo{publisher}{Springer}, pp.
  \bibinfo{pages}{114--129}, \doi{10.1007/3-540-48320-9\_10}.
\newblock \urlprefix\url{https://doi.org/10.1007/3-540-48320-9\_10}.

\bibitemdeclare{inproceedings}{DBLP:conf/coordination/BarbaneraLT20}
\bibitem{DBLP:conf/coordination/BarbaneraLT20}
\bibinfo{author}{Franco \surnamestart Barbanera\surnameend},
  \bibinfo{author}{Ivan \surnamestart Lanese\surnameend} \&
  \bibinfo{author}{Emilio \surnamestart Tuosto\surnameend}
  (\bibinfo{year}{2020}): \emph{\bibinfo{title}{Choreography Automata}}.
\newblock In \bibinfo{editor}{Simon \surnamestart Bliudze\surnameend} \&
  \bibinfo{editor}{Laura \surnamestart Bocchi\surnameend}, editors: {\slshape
  \bibinfo{booktitle}{Coordination Models and Languages - 22nd {IFIP} {WG} 6.1
  International Conference, {COORDINATION} 2020, Held as Part of the 15th
  International Federated Conference on Distributed Computing Techniques,
  DisCoTec 2020, Valletta, Malta, June 15-19, 2020, Proceedings}}, {\slshape
  \bibinfo{series}{Lecture Notes in Computer Science}} \bibinfo{volume}{12134},
  \bibinfo{publisher}{Springer}, pp. \bibinfo{pages}{86--106},
  \doi{10.1007/978-3-030-50029-0\_6}.
\newblock \urlprefix\url{https://doi.org/10.1007/978-3-030-50029-0\_6}.

\bibitemdeclare{inproceedings}{DBLP:conf/tacas/Ben-AbdallahL97}
\bibitem{DBLP:conf/tacas/Ben-AbdallahL97}
\bibinfo{author}{Han{\^{e}}ne \surnamestart Ben{-}Abdallah\surnameend} \&
  \bibinfo{author}{Stefan \surnamestart Leue\surnameend}
  (\bibinfo{year}{1997}): \emph{\bibinfo{title}{Syntactic Detection of Process
  Divergence and Non-local Choice inMessage Sequence Charts}}.
\newblock In \bibinfo{editor}{Ed~\surnamestart Brinksma\surnameend}, editor:
  {\slshape \bibinfo{booktitle}{Tools and Algorithms for Construction and
  Analysis of Systems, Third International Workshop, {TACAS} '97, Enschede, The
  Netherlands, April 2-4, 1997, Proceedings}}, {\slshape
  \bibinfo{series}{Lecture Notes in Computer Science}} \bibinfo{volume}{1217},
  \bibinfo{publisher}{Springer}, pp. \bibinfo{pages}{259--274},
  \doi{10.1007/BFb0035393}.
\newblock \urlprefix\url{https://doi.org/10.1007/BFb0035393}.

\bibitemdeclare{inproceedings}{DBLP:conf/sas/BoigelotGWW97}
\bibitem{DBLP:conf/sas/BoigelotGWW97}
\bibinfo{author}{Bernard \surnamestart Boigelot\surnameend},
  \bibinfo{author}{Patrice \surnamestart Godefroid\surnameend},
  \bibinfo{author}{Bernard \surnamestart Willems\surnameend} \&
  \bibinfo{author}{Pierre \surnamestart Wolper\surnameend}
  (\bibinfo{year}{1997}): \emph{\bibinfo{title}{The Power of QDDs (Extended
  Abstract)}}.
\newblock In \bibinfo{editor}{Pascal~Van \surnamestart Hentenryck\surnameend},
  editor: {\slshape \bibinfo{booktitle}{Static Analysis, 4th International
  Symposium, {SAS} '97, Paris, France, September 8-10, 1997, Proceedings}},
  {\slshape \bibinfo{series}{Lecture Notes in Computer Science}}
  \bibinfo{volume}{1302}, \bibinfo{publisher}{Springer}, pp.
  \bibinfo{pages}{172--186}, \doi{10.1007/BFb0032741}.
\newblock \urlprefix\url{https://doi.org/10.1007/BFb0032741}.

\bibitemdeclare{inproceedings}{DBLP:conf/concur/BolligFS20}
\bibitem{DBLP:conf/concur/BolligFS20}
\bibinfo{author}{Benedikt \surnamestart Bollig\surnameend},
  \bibinfo{author}{Alain \surnamestart Finkel\surnameend} \&
  \bibinfo{author}{Amrita \surnamestart Suresh\surnameend}
  (\bibinfo{year}{2020}): \emph{\bibinfo{title}{Bounded Reachability Problems
  Are Decidable in {FIFO} Machines}}.
\newblock In \bibinfo{editor}{Igor \surnamestart Konnov\surnameend} \&
  \bibinfo{editor}{Laura \surnamestart Kov{\'{a}}cs\surnameend}, editors:
  {\slshape \bibinfo{booktitle}{31st International Conference on Concurrency
  Theory, {CONCUR} 2020, September 1-4, 2020, Vienna, Austria (Virtual
  Conference)}}, {\slshape \bibinfo{series}{LIPIcs}} \bibinfo{volume}{171},
  \bibinfo{publisher}{Schloss Dagstuhl - Leibniz-Zentrum f{\"{u}}r Informatik},
  pp. \bibinfo{pages}{49:1--49:17}, \doi{10.4230/LIPIcs.CONCUR.2020.49}.
\newblock \urlprefix\url{https://doi.org/10.4230/LIPIcs.CONCUR.2020.49}.

\bibitemdeclare{article}{DBLP:journals/corr/abs-1904-06942}
\bibitem{DBLP:journals/corr/abs-1904-06942}
\bibinfo{author}{Benedikt \surnamestart Bollig\surnameend} \&
  \bibinfo{author}{Paul \surnamestart Gastin\surnameend}
  (\bibinfo{year}{2019}): \emph{\bibinfo{title}{Non-Sequential Theory of
  Distributed Systems}}.
\newblock {\slshape \bibinfo{journal}{CoRR}} \bibinfo{volume}{abs/1904.06942}.
\newblock \eprint{1904.06942}.

\bibitemdeclare{inproceedings}{DBLP:conf/concur/BolligGFLLS21}
\bibitem{DBLP:conf/concur/BolligGFLLS21}
\bibinfo{author}{Benedikt \surnamestart Bollig\surnameend},
  \bibinfo{author}{Cinzia~Di \surnamestart Giusto\surnameend},
  \bibinfo{author}{Alain \surnamestart Finkel\surnameend},
  \bibinfo{author}{Laetitia \surnamestart Laversa\surnameend},
  \bibinfo{author}{{\'{E}}tienne \surnamestart Lozes\surnameend} \&
  \bibinfo{author}{Amrita \surnamestart Suresh\surnameend}
  (\bibinfo{year}{2021}): \emph{\bibinfo{title}{A Unifying Framework for
  Deciding Synchronizability}}.
\newblock In \bibinfo{editor}{Serge \surnamestart Haddad\surnameend} \&
  \bibinfo{editor}{Daniele \surnamestart Varacca\surnameend}, editors:
  {\slshape \bibinfo{booktitle}{32nd International Conference on Concurrency
  Theory, {CONCUR} 2021, August 24-27, 2021, Virtual Conference}}, {\slshape
  \bibinfo{series}{LIPIcs}} \bibinfo{volume}{203}, \bibinfo{publisher}{Schloss
  Dagstuhl - Leibniz-Zentrum f{\"{u}}r Informatik}, pp.
  \bibinfo{pages}{14:1--14:18}, \doi{10.4230/LIPIcs.CONCUR.2021.14}.
\newblock \urlprefix\url{https://doi.org/10.4230/LIPIcs.CONCUR.2021.14}.

\bibitemdeclare{inproceedings}{DBLP:conf/cav/BouajjaniEJQ18}
\bibitem{DBLP:conf/cav/BouajjaniEJQ18}
\bibinfo{author}{Ahmed \surnamestart Bouajjani\surnameend},
  \bibinfo{author}{Constantin \surnamestart Enea\surnameend},
  \bibinfo{author}{Kailiang \surnamestart Ji\surnameend} \&
  \bibinfo{author}{Shaz \surnamestart Qadeer\surnameend}
  (\bibinfo{year}{2018}): \emph{\bibinfo{title}{On the Completeness of
  Verifying Message Passing Programs Under Bounded Asynchrony}}.
\newblock In \bibinfo{editor}{Hana \surnamestart Chockler\surnameend} \&
  \bibinfo{editor}{Georg \surnamestart Weissenbacher\surnameend}, editors:
  {\slshape \bibinfo{booktitle}{Computer Aided Verification - 30th
  International Conference, {CAV} 2018, Held as Part of the Federated Logic
  Conference, FloC 2018, Oxford, UK, July 14-17, 2018, Proceedings, Part
  {II}}}, {\slshape \bibinfo{series}{Lecture Notes in Computer Science}}
  \bibinfo{volume}{10982}, \bibinfo{publisher}{Springer}, pp.
  \bibinfo{pages}{372--391}, \doi{10.1007/978-3-319-96142-2\_23}.
\newblock \urlprefix\url{https://doi.org/10.1007/978-3-319-96142-2\_23}.

\bibitemdeclare{article}{DBLP:journals/jacm/BrandZ83}
\bibitem{DBLP:journals/jacm/BrandZ83}
\bibinfo{author}{Daniel \surnamestart Brand\surnameend} \&
  \bibinfo{author}{Pitro \surnamestart Zafiropulo\surnameend}
  (\bibinfo{year}{1983}): \emph{\bibinfo{title}{On Communicating Finite-State
  Machines}}.
\newblock {\slshape \bibinfo{journal}{J. {ACM}}}
  \bibinfo{volume}{30}(\bibinfo{number}{2}), pp. \bibinfo{pages}{323--342},
  \doi{10.1145/322374.322380}.
\newblock \urlprefix\url{https://doi.org/10.1145/322374.322380}.

\bibitemdeclare{inproceedings}{Carbone2005ATB}
\bibitem{Carbone2005ATB}
\bibinfo{author}{Marco \surnamestart Carbone\surnameend},
  \bibinfo{author}{Kohei \surnamestart Honda\surnameend},
  \bibinfo{author}{N.~\surnamestart Yoshida\surnameend},
  \bibinfo{author}{R.~\surnamestart Milner\surnameend},
  \bibinfo{author}{G.~\surnamestart Brown\surnameend} \& \bibinfo{author}{Steve
  \surnamestart Ross-Talbot\surnameend} (\bibinfo{year}{2005}):
  \emph{\bibinfo{title}{A Theoretical Basis of Communication-Centred Concurrent
  Programming}}.

\bibitemdeclare{article}{DBLP:journals/corr/abs-1203-0780}
\bibitem{DBLP:journals/corr/abs-1203-0780}
\bibinfo{author}{Giuseppe \surnamestart Castagna\surnameend},
  \bibinfo{author}{Mariangiola \surnamestart Dezani{-}Ciancaglini\surnameend}
  \& \bibinfo{author}{Luca \surnamestart Padovani\surnameend}
  (\bibinfo{year}{2012}): \emph{\bibinfo{title}{On Global Types and Multi-Party
  Session}}.
\newblock {\slshape \bibinfo{journal}{Log. Methods Comput. Sci.}}
  \bibinfo{volume}{8}(\bibinfo{number}{1}), \doi{10.2168/LMCS-8(1:24)2012}.
\newblock \urlprefix\url{https://doi.org/10.2168/LMCS-8(1:24)2012}.

\bibitemdeclare{article}{DBLP:journals/iandc/CeceF05}
\bibitem{DBLP:journals/iandc/CeceF05}
\bibinfo{author}{G{\'{e}}rard \surnamestart C{\'{e}}c{\'{e}}\surnameend} \&
  \bibinfo{author}{Alain \surnamestart Finkel\surnameend}
  (\bibinfo{year}{2005}): \emph{\bibinfo{title}{Verification of programs with
  half-duplex communication}}.
\newblock {\slshape \bibinfo{journal}{Inf. Comput.}}
  \bibinfo{volume}{202}(\bibinfo{number}{2}), pp. \bibinfo{pages}{166--190},
  \doi{10.1016/j.ic.2005.05.006}.
\newblock \urlprefix\url{https://doi.org/10.1016/j.ic.2005.05.006}.

\bibitemdeclare{inproceedings}{DBLP:conf/sefm/DanHC10}
\bibitem{DBLP:conf/sefm/DanHC10}
\bibinfo{author}{Haitao \surnamestart Dan\surnameend},
  \bibinfo{author}{Robert~M. \surnamestart Hierons\surnameend} \&
  \bibinfo{author}{Steve \surnamestart Counsell\surnameend}
  (\bibinfo{year}{2010}): \emph{\bibinfo{title}{Non-local Choice and Implied
  Scenarios}}.
\newblock In \bibinfo{editor}{Jos{\'{e}}~Luiz \surnamestart
  Fiadeiro\surnameend}, \bibinfo{editor}{Stefania \surnamestart
  Gnesi\surnameend} \& \bibinfo{editor}{Andrea \surnamestart
  Maggiolo{-}Schettini\surnameend}, editors: {\slshape \bibinfo{booktitle}{8th
  {IEEE} International Conference on Software Engineering and Formal Methods,
  {SEFM} 2010, Pisa, Italy, 13-18 September 2010}}, \bibinfo{publisher}{{IEEE}
  Computer Society}, pp. \bibinfo{pages}{53--62}, \doi{10.1109/SEFM.2010.14}.
\newblock \urlprefix\url{https://doi.org/10.1109/SEFM.2010.14}.

\bibitemdeclare{inproceedings}{DBLP:conf/esop/DenielouY12}
\bibitem{DBLP:conf/esop/DenielouY12}
\bibinfo{author}{Pierre{-}Malo \surnamestart Deni{\'{e}}lou\surnameend} \&
  \bibinfo{author}{Nobuko \surnamestart Yoshida\surnameend}
  (\bibinfo{year}{2012}): \emph{\bibinfo{title}{Multiparty Session Types Meet
  Communicating Automata}}.
\newblock In \bibinfo{editor}{Helmut \surnamestart Seidl\surnameend}, editor:
  {\slshape \bibinfo{booktitle}{Programming Languages and Systems - 21st
  European Symposium on Programming, {ESOP} 2012, Held as Part of the European
  Joint Conferences on Theory and Practice of Software, {ETAPS} 2012, Tallinn,
  Estonia, March 24 - April 1, 2012. Proceedings}}, {\slshape
  \bibinfo{series}{Lecture Notes in Computer Science}} \bibinfo{volume}{7211},
  \bibinfo{publisher}{Springer}, pp. \bibinfo{pages}{194--213},
  \doi{10.1007/978-3-642-28869-2\_10}.
\newblock \urlprefix\url{https://doi.org/10.1007/978-3-642-28869-2\_10}.

\bibitemdeclare{inproceedings}{DBLP:conf/eurosys/FahndrichAHHHLL06}
\bibitem{DBLP:conf/eurosys/FahndrichAHHHLL06}
\bibinfo{author}{Manuel \surnamestart F{\"{a}}hndrich\surnameend},
  \bibinfo{author}{Mark \surnamestart Aiken\surnameend}, \bibinfo{author}{Chris
  \surnamestart Hawblitzel\surnameend}, \bibinfo{author}{Orion \surnamestart
  Hodson\surnameend}, \bibinfo{author}{Galen~C. \surnamestart Hunt\surnameend},
  \bibinfo{author}{James~R. \surnamestart Larus\surnameend} \&
  \bibinfo{author}{Steven \surnamestart Levi\surnameend}
  (\bibinfo{year}{2006}): \emph{\bibinfo{title}{Language support for fast and
  reliable message-based communication in singularity {OS}}}.
\newblock In \bibinfo{editor}{Yolande \surnamestart Berbers\surnameend} \&
  \bibinfo{editor}{Willy \surnamestart Zwaenepoel\surnameend}, editors:
  {\slshape \bibinfo{booktitle}{Proceedings of the 2006 EuroSys Conference,
  Leuven, Belgium, April 18-21, 2006}}, \bibinfo{publisher}{{ACM}}, pp.
  \bibinfo{pages}{177--190}, \doi{10.1145/1217935.1217953}.
\newblock \urlprefix\url{https://doi.org/10.1145/1217935.1217953}.

\bibitemdeclare{inproceedings}{DBLP:conf/icalp/FinkelL17}
\bibitem{DBLP:conf/icalp/FinkelL17}
\bibinfo{author}{Alain \surnamestart Finkel\surnameend} \&
  \bibinfo{author}{{\'{E}}tienne \surnamestart Lozes\surnameend}
  (\bibinfo{year}{2017}): \emph{\bibinfo{title}{Synchronizability of
  Communicating Finite State Machines is not Decidable}}.
\newblock In \bibinfo{editor}{Ioannis \surnamestart
  Chatzigiannakis\surnameend}, \bibinfo{editor}{Piotr \surnamestart
  Indyk\surnameend}, \bibinfo{editor}{Fabian \surnamestart Kuhn\surnameend} \&
  \bibinfo{editor}{Anca \surnamestart Muscholl\surnameend}, editors: {\slshape
  \bibinfo{booktitle}{44th International Colloquium on Automata, Languages, and
  Programming, {ICALP} 2017, July 10-14, 2017, Warsaw, Poland}}, {\slshape
  \bibinfo{series}{LIPIcs}}~\bibinfo{volume}{80}, \bibinfo{publisher}{Schloss
  Dagstuhl - Leibniz-Zentrum f{\"{u}}r Informatik}, pp.
  \bibinfo{pages}{122:1--122:14}, \doi{10.4230/LIPIcs.ICALP.2017.122}.
\newblock \urlprefix\url{https://doi.org/10.4230/LIPIcs.ICALP.2017.122}.

\bibitemdeclare{article}{DBLP:journals/jfp/GayV10}
\bibitem{DBLP:journals/jfp/GayV10}
\bibinfo{author}{Simon~J. \surnamestart Gay\surnameend} \&
  \bibinfo{author}{Vasco~Thudichum \surnamestart Vasconcelos\surnameend}
  (\bibinfo{year}{2010}): \emph{\bibinfo{title}{Linear type theory for
  asynchronous session types}}.
\newblock {\slshape \bibinfo{journal}{J. Funct. Program.}}
  \bibinfo{volume}{20}(\bibinfo{number}{1}), pp. \bibinfo{pages}{19--50},
  \doi{10.1017/S0956796809990268}.
\newblock \urlprefix\url{https://doi.org/10.1017/S0956796809990268}.

\bibitemdeclare{inproceedings}{DBLP:conf/concur/GazagnaireGHTY07}
\bibitem{DBLP:conf/concur/GazagnaireGHTY07}
\bibinfo{author}{Thomas \surnamestart Gazagnaire\surnameend},
  \bibinfo{author}{Blaise \surnamestart Genest\surnameend},
  \bibinfo{author}{Lo{\"{\i}}c \surnamestart H{\'{e}}lou{\"{e}}t\surnameend},
  \bibinfo{author}{P.~S. \surnamestart Thiagarajan\surnameend} \&
  \bibinfo{author}{Shaofa \surnamestart Yang\surnameend}
  (\bibinfo{year}{2007}): \emph{\bibinfo{title}{Causal Message Sequence
  Charts}}.
\newblock In \bibinfo{editor}{Lu{\'{\i}}s \surnamestart Caires\surnameend} \&
  \bibinfo{editor}{Vasco~Thudichum \surnamestart Vasconcelos\surnameend},
  editors: {\slshape \bibinfo{booktitle}{{CONCUR} 2007 - Concurrency Theory,
  18th International Conference, {CONCUR} 2007, Lisbon, Portugal, September
  3-8, 2007, Proceedings}}, {\slshape \bibinfo{series}{Lecture Notes in
  Computer Science}} \bibinfo{volume}{4703}, \bibinfo{publisher}{Springer}, pp.
  \bibinfo{pages}{166--180}, \doi{10.1007/978-3-540-74407-8\_12}.
\newblock \urlprefix\url{https://doi.org/10.1007/978-3-540-74407-8\_12}.

\bibitemdeclare{article}{DBLP:journals/fuin/GenestKM07}
\bibitem{DBLP:journals/fuin/GenestKM07}
\bibinfo{author}{Blaise \surnamestart Genest\surnameend},
  \bibinfo{author}{Dietrich \surnamestart Kuske\surnameend} \&
  \bibinfo{author}{Anca \surnamestart Muscholl\surnameend}
  (\bibinfo{year}{2007}): \emph{\bibinfo{title}{On Communicating Automata with
  Bounded Channels}}.
\newblock {\slshape \bibinfo{journal}{Fundam. Inform.}}
  \bibinfo{volume}{80}(\bibinfo{number}{1-3}), pp. \bibinfo{pages}{147--167}.
\newblock
  \urlprefix\url{http://content.iospress.com/articles/fundamenta-informaticae/fi80-1-3-09}.

\bibitemdeclare{inproceedings}{DBLP:conf/acsd/GenestM05}
\bibitem{DBLP:conf/acsd/GenestM05}
\bibinfo{author}{Blaise \surnamestart Genest\surnameend} \&
  \bibinfo{author}{Anca \surnamestart Muscholl\surnameend}
  (\bibinfo{year}{2005}): \emph{\bibinfo{title}{Message Sequence Charts: {A}
  Survey}}.
\newblock In: {\slshape \bibinfo{booktitle}{Fifth International Conference on
  Application of Concurrency to System Design {(ACSD} 2005), 6-9 June 2005, St.
  Malo, France}}, \bibinfo{publisher}{{IEEE} Computer Society}, pp.
  \bibinfo{pages}{2--4}, \doi{10.1109/ACSD.2005.25}.
\newblock \urlprefix\url{https://doi.org/10.1109/ACSD.2005.25}.

\bibitemdeclare{inproceedings}{DBLP:conf/ac/GenestMP03}
\bibitem{DBLP:conf/ac/GenestMP03}
\bibinfo{author}{Blaise \surnamestart Genest\surnameend}, \bibinfo{author}{Anca
  \surnamestart Muscholl\surnameend} \& \bibinfo{author}{Doron~A. \surnamestart
  Peled\surnameend} (\bibinfo{year}{2003}): \emph{\bibinfo{title}{Message
  Sequence Charts}}.
\newblock In \bibinfo{editor}{J{\"{o}}rg \surnamestart Desel\surnameend},
  \bibinfo{editor}{Wolfgang \surnamestart Reisig\surnameend} \&
  \bibinfo{editor}{Grzegorz \surnamestart Rozenberg\surnameend}, editors:
  {\slshape \bibinfo{booktitle}{Lectures on Concurrency and Petri Nets,
  Advances in Petri Nets [This tutorial volume originates from the 4th Advanced
  Course on Petri Nets, {ACPN} 2003, held in Eichst{\"{a}}tt, Germany in
  September 2003. In addition to lectures given at {ACPN} 2003, additional
  chapters have been commissioned]}}, {\slshape \bibinfo{series}{Lecture Notes
  in Computer Science}} \bibinfo{volume}{3098}, \bibinfo{publisher}{Springer},
  pp. \bibinfo{pages}{537--558}, \doi{10.1007/978-3-540-27755-2\_15}.
\newblock \urlprefix\url{https://doi.org/10.1007/978-3-540-27755-2\_15}.

\bibitemdeclare{article}{DBLP:journals/jcss/GenestMSZ06}
\bibitem{DBLP:journals/jcss/GenestMSZ06}
\bibinfo{author}{Blaise \surnamestart Genest\surnameend}, \bibinfo{author}{Anca
  \surnamestart Muscholl\surnameend}, \bibinfo{author}{Helmut \surnamestart
  Seidl\surnameend} \& \bibinfo{author}{Marc \surnamestart Zeitoun\surnameend}
  (\bibinfo{year}{2006}): \emph{\bibinfo{title}{Infinite-state high-level MSCs:
  Model-checking and realizability}}.
\newblock {\slshape \bibinfo{journal}{J. Comput. Syst. Sci.}}
  \bibinfo{volume}{72}(\bibinfo{number}{4}), pp. \bibinfo{pages}{617--647},
  \doi{10.1016/j.jcss.2005.09.007}.
\newblock \urlprefix\url{https://doi.org/10.1016/j.jcss.2005.09.007}.

\bibitemdeclare{inproceedings}{DBLP:conf/fossacs/GiustoLL20}
\bibitem{DBLP:conf/fossacs/GiustoLL20}
\bibinfo{author}{Cinzia~Di \surnamestart Giusto\surnameend},
  \bibinfo{author}{Laetitia \surnamestart Laversa\surnameend} \&
  \bibinfo{author}{{\'{E}}tienne \surnamestart Lozes\surnameend}
  (\bibinfo{year}{2020}): \emph{\bibinfo{title}{On the k-synchronizability of
  Systems}}.
\newblock In \bibinfo{editor}{Jean \surnamestart Goubault{-}Larrecq\surnameend}
  \& \bibinfo{editor}{Barbara \surnamestart K{\"{o}}nig\surnameend}, editors:
  {\slshape \bibinfo{booktitle}{Foundations of Software Science and Computation
  Structures - 23rd International Conference, {FOSSACS} 2020, Held as Part of
  the European Joint Conferences on Theory and Practice of Software, {ETAPS}
  2020, Dublin, Ireland, April 25-30, 2020, Proceedings}}, {\slshape
  \bibinfo{series}{Lecture Notes in Computer Science}} \bibinfo{volume}{12077},
  \bibinfo{publisher}{Springer}, pp. \bibinfo{pages}{157--176},
  \doi{10.1007/978-3-030-45231-5\_9}.
\newblock \urlprefix\url{https://doi.org/10.1007/978-3-030-45231-5\_9}.

\bibitemdeclare{inproceedings}{NO-DBLP-wrong-local-choice}
\bibitem{NO-DBLP-wrong-local-choice}
\bibinfo{author}{Lo{\"{\i}}c \surnamestart H{\'{e}}lou{\"{e}}t\surnameend} \&
  \bibinfo{author}{Claude \surnamestart Jard\surnameend}
  (\bibinfo{year}{2000}): \emph{\bibinfo{title}{Conditions for synthesis of
  communicating automata from {HMSCs}}}.
\newblock In: {\slshape \bibinfo{booktitle}{In 5th International Workshop on
  Formal Methods for Industrial Critical Systems (FMICS)}}.

\bibitemdeclare{article}{DBLP:journals/corr/abs-1209-0359}
\bibitem{DBLP:journals/corr/abs-1209-0359}
\bibinfo{author}{Alexander \surnamestart Heu{\ss}ner\surnameend},
  \bibinfo{author}{J{\'{e}}r{\^{o}}me \surnamestart Leroux\surnameend},
  \bibinfo{author}{Anca \surnamestart Muscholl\surnameend} \&
  \bibinfo{author}{Gr{\'{e}}goire \surnamestart Sutre\surnameend}
  (\bibinfo{year}{2012}): \emph{\bibinfo{title}{Reachability Analysis of
  Communicating Pushdown Systems}}.
\newblock {\slshape \bibinfo{journal}{Log. Methods Comput. Sci.}}
  \bibinfo{volume}{8}(\bibinfo{number}{3}), \doi{10.2168/LMCS-8(3:23)2012}.
\newblock \urlprefix\url{https://doi.org/10.2168/LMCS-8(3:23)2012}.

\bibitemdeclare{inproceedings}{DBLP:conf/concur/Honda93}
\bibitem{DBLP:conf/concur/Honda93}
\bibinfo{author}{Kohei \surnamestart Honda\surnameend} (\bibinfo{year}{1993}):
  \emph{\bibinfo{title}{Types for Dyadic Interaction}}.
\newblock In \bibinfo{editor}{Eike \surnamestart Best\surnameend}, editor:
  {\slshape \bibinfo{booktitle}{{CONCUR} '93, 4th International Conference on
  Concurrency Theory, Hildesheim, Germany, August 23-26, 1993, Proceedings}},
  {\slshape \bibinfo{series}{Lecture Notes in Computer Science}}
  \bibinfo{volume}{715}, \bibinfo{publisher}{Springer}, pp.
  \bibinfo{pages}{509--523}, \doi{10.1007/3-540-57208-2\_35}.
\newblock \urlprefix\url{https://doi.org/10.1007/3-540-57208-2\_35}.

\bibitemdeclare{inproceedings}{DBLP:conf/popl/HondaYC08}
\bibitem{DBLP:conf/popl/HondaYC08}
\bibinfo{author}{Kohei \surnamestart Honda\surnameend}, \bibinfo{author}{Nobuko
  \surnamestart Yoshida\surnameend} \& \bibinfo{author}{Marco \surnamestart
  Carbone\surnameend} (\bibinfo{year}{2008}): \emph{\bibinfo{title}{Multiparty
  asynchronous session types}}.
\newblock In \bibinfo{editor}{George~C. \surnamestart Necula\surnameend} \&
  \bibinfo{editor}{Philip \surnamestart Wadler\surnameend}, editors: {\slshape
  \bibinfo{booktitle}{Proceedings of the 35th {ACM} {SIGPLAN-SIGACT} Symposium
  on Principles of Programming Languages, {POPL} 2008, San Francisco,
  California, USA, January 7-12, 2008}}, \bibinfo{publisher}{{ACM}}, pp.
  \bibinfo{pages}{273--284}, \doi{10.1145/1328438.1328472}.
\newblock \urlprefix\url{https://doi.org/10.1145/1328438.1328472}.

\bibitemdeclare{article}{DBLP:journals/jacm/HondaYC16}
\bibitem{DBLP:journals/jacm/HondaYC16}
\bibinfo{author}{Kohei \surnamestart Honda\surnameend}, \bibinfo{author}{Nobuko
  \surnamestart Yoshida\surnameend} \& \bibinfo{author}{Marco \surnamestart
  Carbone\surnameend} (\bibinfo{year}{2016}): \emph{\bibinfo{title}{Multiparty
  Asynchronous Session Types}}.
\newblock {\slshape \bibinfo{journal}{J. {ACM}}}
  \bibinfo{volume}{63}(\bibinfo{number}{1}), pp. \bibinfo{pages}{9:1--9:67},
  \doi{10.1145/2827695}.
\newblock \urlprefix\url{https://doi.org/10.1145/2827695}.

\bibitemdeclare{article}{DBLP:journals/mst/Kocher21}
\bibitem{DBLP:journals/mst/Kocher21}
\bibinfo{author}{Chris \surnamestart K{\"{o}}cher\surnameend}
  (\bibinfo{year}{2021}): \emph{\bibinfo{title}{Reachability Problems on
  Reliable and Lossy Queue Automata}}.
\newblock {\slshape \bibinfo{journal}{Theory Comput. Syst.}}
  \bibinfo{volume}{65}(\bibinfo{number}{8}), pp. \bibinfo{pages}{1211--1242},
  \doi{10.1007/s00224-021-10031-2}.
\newblock \urlprefix\url{https://doi.org/10.1007/s00224-021-10031-2}.

\bibitemdeclare{inproceedings}{DBLP:conf/cav/LangeY19}
\bibitem{DBLP:conf/cav/LangeY19}
\bibinfo{author}{Julien \surnamestart Lange\surnameend} \&
  \bibinfo{author}{Nobuko \surnamestart Yoshida\surnameend}
  (\bibinfo{year}{2019}): \emph{\bibinfo{title}{Verifying Asynchronous
  Interactions via Communicating Session Automata}}.
\newblock In \bibinfo{editor}{Isil \surnamestart Dillig\surnameend} \&
  \bibinfo{editor}{Serdar \surnamestart Tasiran\surnameend}, editors: {\slshape
  \bibinfo{booktitle}{Computer Aided Verification - 31st International
  Conference, {CAV} 2019, New York City, NY, USA, July 15-18, 2019,
  Proceedings, Part {I}}}, {\slshape \bibinfo{series}{Lecture Notes in Computer
  Science}} \bibinfo{volume}{11561}, \bibinfo{publisher}{Springer}, pp.
  \bibinfo{pages}{97--117}, \doi{10.1007/978-3-030-25540-4\_6}.
\newblock \urlprefix\url{https://doi.org/10.1007/978-3-030-25540-4\_6}.

\bibitemdeclare{article}{DBLP:journals/tcs/Lohrey03}
\bibitem{DBLP:journals/tcs/Lohrey03}
\bibinfo{author}{Markus \surnamestart Lohrey\surnameend}
  (\bibinfo{year}{2003}): \emph{\bibinfo{title}{Realizability of high-level
  message sequence charts: closing the gaps}}.
\newblock {\slshape \bibinfo{journal}{Theor. Comput. Sci.}}
  \bibinfo{volume}{309}(\bibinfo{number}{1-3}), pp. \bibinfo{pages}{529--554},
  \doi{10.1016/j.tcs.2003.08.002}.
\newblock \urlprefix\url{https://doi.org/10.1016/j.tcs.2003.08.002}.

\bibitemdeclare{inproceedings}{DBLP:conf/icalp/Madhusudan01}
\bibitem{DBLP:conf/icalp/Madhusudan01}
\bibinfo{author}{P.~\surnamestart Madhusudan\surnameend}
  (\bibinfo{year}{2001}): \emph{\bibinfo{title}{Reasoning about Sequential and
  Branching Behaviours of Message Sequence Graphs}}.
\newblock In \bibinfo{editor}{Fernando \surnamestart Orejas\surnameend},
  \bibinfo{editor}{Paul~G. \surnamestart Spirakis\surnameend} \&
  \bibinfo{editor}{Jan \surnamestart van Leeuwen\surnameend}, editors:
  {\slshape \bibinfo{booktitle}{Automata, Languages and Programming, 28th
  International Colloquium, {ICALP} 2001, Crete, Greece, July 8-12, 2001,
  Proceedings}}, {\slshape \bibinfo{series}{Lecture Notes in Computer Science}}
  \bibinfo{volume}{2076}, \bibinfo{publisher}{Springer}, pp.
  \bibinfo{pages}{809--820}, \doi{10.1007/3-540-48224-5\_66}.
\newblock \urlprefix\url{https://doi.org/10.1007/3-540-48224-5\_66}.

\bibitemdeclare{inproceedings}{DBLP:conf/concur/MajumdarMSZ21}
\bibitem{DBLP:conf/concur/MajumdarMSZ21}
\bibinfo{author}{Rupak \surnamestart Majumdar\surnameend},
  \bibinfo{author}{Madhavan \surnamestart Mukund\surnameend},
  \bibinfo{author}{Felix \surnamestart Stutz\surnameend} \&
  \bibinfo{author}{Damien \surnamestart Zufferey\surnameend}
  (\bibinfo{year}{2021}): \emph{\bibinfo{title}{Generalising Projection in
  Asynchronous Multiparty Session Types}}.
\newblock In \bibinfo{editor}{Serge \surnamestart Haddad\surnameend} \&
  \bibinfo{editor}{Daniele \surnamestart Varacca\surnameend}, editors:
  {\slshape \bibinfo{booktitle}{32nd International Conference on Concurrency
  Theory, {CONCUR} 2021, August 24-27, 2021, Virtual Conference}}, {\slshape
  \bibinfo{series}{LIPIcs}} \bibinfo{volume}{203}, \bibinfo{publisher}{Schloss
  Dagstuhl - Leibniz-Zentrum f{\"{u}}r Informatik}, pp.
  \bibinfo{pages}{35:1--35:24}, \doi{10.4230/LIPIcs.CONCUR.2021.35}.
\newblock \urlprefix\url{https://doi.org/10.4230/LIPIcs.CONCUR.2021.35}.

\bibitemdeclare{inproceedings}{DBLP:conf/sdl/MauwR97}
\bibitem{DBLP:conf/sdl/MauwR97}
\bibinfo{author}{Sjouke \surnamestart Mauw\surnameend} \&
  \bibinfo{author}{Michel~A. \surnamestart Reniers\surnameend}
  (\bibinfo{year}{1997}): \emph{\bibinfo{title}{High-level message sequence
  charts}}.
\newblock In \bibinfo{editor}{Ana~R. \surnamestart Cavalli\surnameend} \&
  \bibinfo{editor}{Amardeo \surnamestart Sarma\surnameend}, editors: {\slshape
  \bibinfo{booktitle}{{SDL} '97 Time for Testing, SDL, {MSC} and Trends - 8th
  International {SDL} Forum, Evry, France, 23-29 September 1997, Proceedings}},
  \bibinfo{publisher}{Elsevier}, pp. \bibinfo{pages}{291--306}.

\bibitemdeclare{book}{DBLP:books/daglib/0098267}
\bibitem{DBLP:books/daglib/0098267}
\bibinfo{author}{Robin \surnamestart Milner\surnameend} (\bibinfo{year}{1999}):
  \emph{\bibinfo{title}{Communicating and mobile systems - the Pi-calculus}}.
\newblock \bibinfo{publisher}{Cambridge University Press}.

\bibitemdeclare{inproceedings}{DBLP:conf/fase/MooijGR05}
\bibitem{DBLP:conf/fase/MooijGR05}
\bibinfo{author}{Arjan~J. \surnamestart Mooij\surnameend},
  \bibinfo{author}{Nicolae \surnamestart Goga\surnameend} \&
  \bibinfo{author}{Judi \surnamestart Romijn\surnameend}
  (\bibinfo{year}{2005}): \emph{\bibinfo{title}{Non-local Choice and Beyond:
  Intricacies of {MSC} Choice Nodes}}.
\newblock In \bibinfo{editor}{Maura \surnamestart Cerioli\surnameend}, editor:
  {\slshape \bibinfo{booktitle}{Fundamental Approaches to Software Engineering,
  8th International Conference, {FASE} 2005, Held as Part of the Joint European
  Conferences on Theory and Practice of Software, {ETAPS} 2005, Edinburgh, UK,
  April 4-8, 2005, Proceedings}}, {\slshape \bibinfo{series}{Lecture Notes in
  Computer Science}} \bibinfo{volume}{3442}, \bibinfo{publisher}{Springer}, pp.
  \bibinfo{pages}{273--288}, \doi{10.1007/978-3-540-31984-9\_21}.
\newblock \urlprefix\url{https://doi.org/10.1007/978-3-540-31984-9\_21}.

\bibitemdeclare{inproceedings}{DBLP:conf/fase/Muccini03}
\bibitem{DBLP:conf/fase/Muccini03}
\bibinfo{author}{Henry \surnamestart Muccini\surnameend}
  (\bibinfo{year}{2003}): \emph{\bibinfo{title}{Detecting Implied Scenarios
  Analyzing Non-local Branching Choices}}.
\newblock In \bibinfo{editor}{Mauro \surnamestart Pezz{\`{e}}\surnameend},
  editor: {\slshape \bibinfo{booktitle}{Fundamental Approaches to Software
  Engineering, 6th International Conference, {FASE} 2003, Held as Part of the
  Joint European Conferences on Theory and Practice of Software, {ETAPS} 2003,
  Warsaw, Poland, April 7-11, 2003, Proceedings}}, {\slshape
  \bibinfo{series}{Lecture Notes in Computer Science}} \bibinfo{volume}{2621},
  \bibinfo{publisher}{Springer}, pp. \bibinfo{pages}{372--386},
  \doi{10.1007/3-540-36578-8\_26}.
\newblock \urlprefix\url{https://doi.org/10.1007/3-540-36578-8\_26}.

\bibitemdeclare{article}{DBLP:journals/acta/PengP92}
\bibitem{DBLP:journals/acta/PengP92}
\bibinfo{author}{Wuxu \surnamestart Peng\surnameend} \&
  \bibinfo{author}{S.~\surnamestart Purushothaman\surnameend}
  (\bibinfo{year}{1992}): \emph{\bibinfo{title}{Analysis of a Class of
  Communicating Finite State Machines}}.
\newblock {\slshape \bibinfo{journal}{Acta Informatica}}
  \bibinfo{volume}{29}(\bibinfo{number}{6/7}), pp. \bibinfo{pages}{499--522},
  \doi{10.1007/BF01185558}.
\newblock \urlprefix\url{https://doi.org/10.1007/BF01185558}.

\bibitemdeclare{article}{DBLP:journals/tosem/RoychoudhuryGS12}
\bibitem{DBLP:journals/tosem/RoychoudhuryGS12}
\bibinfo{author}{Abhik \surnamestart Roychoudhury\surnameend},
  \bibinfo{author}{Ankit \surnamestart Goel\surnameend} \&
  \bibinfo{author}{Bikram \surnamestart Sengupta\surnameend}
  (\bibinfo{year}{2012}): \emph{\bibinfo{title}{Symbolic Message Sequence
  Charts}}.
\newblock {\slshape \bibinfo{journal}{{ACM} Trans. Softw. Eng. Methodol.}}
  \bibinfo{volume}{21}(\bibinfo{number}{2}), pp. \bibinfo{pages}{12:1--12:44},
  \doi{10.1145/2089116.2089122}.
\newblock \urlprefix\url{https://doi.org/10.1145/2089116.2089122}.

\bibitemdeclare{article}{DBLP:journals/pacmpl/ScalasY19}
\bibitem{DBLP:journals/pacmpl/ScalasY19}
\bibinfo{author}{Alceste \surnamestart Scalas\surnameend} \&
  \bibinfo{author}{Nobuko \surnamestart Yoshida\surnameend}
  (\bibinfo{year}{2019}): \emph{\bibinfo{title}{Less is more: multiparty
  session types revisited}}.
\newblock {\slshape \bibinfo{journal}{Proc. {ACM} Program. Lang.}}
  \bibinfo{volume}{3}(\bibinfo{number}{{POPL}}), pp.
  \bibinfo{pages}{30:1--30:29}, \doi{10.1145/3290343}.
\newblock \urlprefix\url{https://doi.org/10.1145/3290343}.

\bibitemdeclare{inproceedings}{DBLP:conf/tacas/TorreMP08}
\bibitem{DBLP:conf/tacas/TorreMP08}
\bibinfo{author}{Salvatore~La \surnamestart Torre\surnameend},
  \bibinfo{author}{P.~\surnamestart Madhusudan\surnameend} \&
  \bibinfo{author}{Gennaro \surnamestart Parlato\surnameend}
  (\bibinfo{year}{2008}): \emph{\bibinfo{title}{Context-Bounded Analysis of
  Concurrent Queue Systems}}.
\newblock In \bibinfo{editor}{C.~R. \surnamestart Ramakrishnan\surnameend} \&
  \bibinfo{editor}{Jakob \surnamestart Rehof\surnameend}, editors: {\slshape
  \bibinfo{booktitle}{Tools and Algorithms for the Construction and Analysis of
  Systems, 14th International Conference, {TACAS} 2008, Held as Part of the
  Joint European Conferences on Theory and Practice of Software, {ETAPS} 2008,
  Budapest, Hungary, March 29-April 6, 2008. Proceedings}}, {\slshape
  \bibinfo{series}{Lecture Notes in Computer Science}} \bibinfo{volume}{4963},
  \bibinfo{publisher}{Springer}, pp. \bibinfo{pages}{299--314},
  \doi{10.1007/978-3-540-78800-3\_21}.
\newblock \urlprefix\url{https://doi.org/10.1007/978-3-540-78800-3\_21}.

\bibitemdeclare{article}{DBLP:journals/tcs/TouiliA10}
\bibitem{DBLP:journals/tcs/TouiliA10}
\bibinfo{author}{Tayssir \surnamestart Touili\surnameend} \&
  \bibinfo{author}{Mohamed~Faouzi \surnamestart Atig\surnameend}
  (\bibinfo{year}{2010}): \emph{\bibinfo{title}{Verifying parallel programs
  with dynamic communication structures}}.
\newblock {\slshape \bibinfo{journal}{Theor. Comput. Sci.}}
  \bibinfo{volume}{411}(\bibinfo{number}{38-39}), pp.
  \bibinfo{pages}{3460--3468}, \doi{10.1016/j.tcs.2010.05.028}.
\newblock \urlprefix\url{https://doi.org/10.1016/j.tcs.2010.05.028}.

\bibitemdeclare{techreport}{z120-standard}
\bibitem{z120-standard}
\bibinfo{author}{International~Telecommunication \surnamestart
  Union\surnameend} (\bibinfo{year}{1996}): \emph{\bibinfo{title}{Z.120:
  Message Sequence Chart}}.
\newblock \bibinfo{type}{Technical Report}, \bibinfo{institution}{International
  Telecommunication Union}.
\newblock \urlprefix\url{https://www.itu.int/rec/T-REC-Z.120}.

\end{thebibliography}

\clearpage
\appendix
\section{Standard Definitions}
\label{app:standard-defs}

\subsection{Notations for Prefixes}
\label{app:prefixes}

In this section, we provide more notation for prefixes that is only used in the Appendix.

\begin{definition}
Let $w$ be some word.
The set $\pref(w)$ contains all prefixes of $w$.
We lift $\pref(\hole)$ to languages as expected:
$\pref(L) = \set{u \mid \exists v\in L.\; u \leq v}$.
\end{definition}

\subsection{Message Sequence Charts}
\label{app:mscs}

\begin{definition}[Concatenation of MSCs]
\label{def:concatenation-msc}
Let $M_i = (\eventnodes_i, p_i, f_i, l_i, (\leq^i_p)_{\procA\in𝓟})$ for $i \in \set{1,2}$ where $M_1$ is a BMSC and $M_2$ is an MSC
with disjoint sets of events, i.e., $\eventnodes_1\cap \eventnodes_2 = \emptyset$.
We define their \emph{concatenation} $M_1\cdot M_2$ as the MSC
$M = (\eventnodes, p, f, l, (\leq_p)_{\procA\in𝓟})$ where:
\begin{itemize}
 \item $\eventnodes
 \quad \is \quad
 \eventnodes₁ \; \union \; \eventnodes₂$,
 \item $
            \text{for } 
            \zeta \in \set{p, f, l}: 
            \quad
            \zeta(e) \is
            \begin{cases}
              \zeta(e)         & \text{if } e \in \eventnodes₁ \\
              \zeta(e)         & \text{if } e \in \eventnodes₂
            \end{cases} 
       $, and
 \item $\forall \procA \in 𝓟: \quad$
$
                \leq_{\procA}
                    \quad \is \quad
                \leq^1_{\procA}
                \; \union \;
                \leq^2_{\procA}
                \; \union \;
                \set{(e₁,e₂) \mid  \, e₁\in \eventnodes₁ \, \land \, e₂\in \eventnodes₂ \, \land  p(e_1) = p(e_2)=\procA }.
       $
\end{itemize}
\end{definition}
Note that we did not define the most general concatenation operator.
First, one could also concatenate prefix MSCs and second we could have allowed an infinite set of events for $\procA$ in $M_1$ if the set of events for $\procA$ is empty in $M_2$, however, this definition suits our purposes.

\subsection{Semantics of HMSCs}
\label{app:semantics-hmsc}

\begin{definition}[Language of an HMSC]
\label{def:lang-hmsc}
Let $H = (V, \edges, \vertexA^I\negmedspace, V^T\negmedspace\!, \mu)$ be an HMSC.
The language of $H$ is defined as
\begin{small}
\begin{align*}
\lang(H) \is
& \;
\set{
    w
    \mid 
    w \in \lang(\mu(\vertexA_1) \mu(\vertexA_2) \ldots \mu(\vertexA_n))
    \text{ with }
    \vertexA_1 = v^I
    \land
    \forall \, 0 \leq i < n: \,
    (v_i, v_{i+1}) \in \edges
    \land
    v_n \in V^T
}
\\
&
\;
\union \;
\set{
    w \mid
    w \in \lang(\mu(\vertexA_1) \mu(\vertexA_2) \ldots )
    \text{ with }
    \vertexA_1 = v^I
    \land
    \forall \, i \geq 0: \,
    (v_i, v_{i+1}) \in \edges
}
\end{align*}
\end{small}

\end{definition}

\subsection{Semantics of CSMs}
\label{app:semantics-csm}

With $\channels = \set{\channel{\procA}{\procB} \mid \procA,\procB\in \Procs, \procA\neq \procB}$, we denote the set of channels.
The set of global states of a CSM is given by $\prod_{\procA \in \Procs} Q_\procA$.
Given a global state $q$, $q_\procA$ denotes the state of $\procA$ in $q$.
A~\emph{configuration} of a CSM $\mathcal{A}$ is a pair $(q, \xi)$, where $q$ is a global state and
$\xi : \channels \rightarrow \MsgVals^\infty$ is a mapping of each channel to its current content.
The initial configuration $(q_0, \xi_\emptystring)$ consists of  a global state $q_0$ where the state of each process is the initial state  $q_{0,\procA}$ of $A_\procA$ and a mapping
$\xi_\emptystring$ which maps each channel to the empty word~$\emptystring$.
A~configuration $(q, \xi)$ is said to be \emph{final} iff each individual local state $q_\procA$ is final for every $\procA$ and $\xi$ is $\xi_{\emptystring}$.

The global transition relation $\rightarrow$ is defined as follows:
\begin{itemize}
\item
$(q,\xi) \xrightarrow{\snd{\procA}{\procB}{\val}} (q',\xi')$ if
$(q_\procA, \snd{\procA}{\procB}{\val}, q'_\procA)\in\delta_\procA$,
$q_\procC = q'_\procC$ for every process $\procC \neq \procA$,
$\xi'(\channel{\procA}{\procB}) =  \xi(\channel{\procA}{\procB})\cdot\val$ and $\xi'(c) = \xi(c)$ for every other channel $c\in \channels$.

\item
$(q,\xi) \xrightarrow{\rcv{\procA}{\procB}{\val}} (q',\xi')$ if
$(q_\procB, \rcv{\procA}{\procB}{\val}, q'_\procB)\in\delta_\procB$,
$q_\procC = q'_\procC$ for every process $\procC \neq \procB$,
$\xi(\channel{\procA}{\procB}) = \val\cdot \xi'(\channel{\procA}{\procB})$
and $\xi'(c) = \xi(c)$ for every other channel $c\in \channels$.

\item
$(q,\xi) \xrightarrow{\emptystring} (q',\xi)$ if
$(q_\procA, \emptystring, q'_\procA)\in\delta_\procA$ for some process
$\procA$, and
$q_\procB = q'_\procB$ for every process $\procB \neq \procA$.
\end{itemize}

\noindent
We might call a run of $\CSM{A}$ (which might not end in a final state) an \emph{execution prefix} of~$\CSM{A}$.

A run of the CSM always starts with an initial configuration $(q_0, \xi_0)$, and is a finite or infinite sequence
$(q_0, \xi_0) \xrightarrow{w_0} (q_1, \xi_1) \xrightarrow{w_1} \ldots$
for which  $(q_i,\xi_i) \xrightarrow{w_i} (q_{i+1},\xi_{i+1})$.
The word $w_0w_1 \ldots \in\Sigma^\infty$ is said to be the \emph{trace} of the run.
A run is called maximal if it is either infinite or finite and ends in a final configuration. As before, the trace of a maximal run is maximal.
The language $\lang(\mathcal{A})$ of the CSM $\mathcal{A}$ consists of its set of maximal traces.

 \section{Supplementary Material on Channel Restrictions for \cref{sec:channel-restrictions}}

\subsection{Equivalence of Definitions for Half-duplex Communication}
\label{app:half-duplex}

In this section, we prove that our definition of half-duplex (\cref{def:half-duplex}) and the original definition by C{\'{e}}c{\'{e}} and Finkel are equivalent.
To be precise, we consider the \emph{natural generalisation} \cite[Section~4]{DBLP:journals/iandc/CeceF05} of the half-duplex definition
\cite[Def.~8]{DBLP:journals/iandc/CeceF05}.

\begin{definition}[Natural generalisation of half-duplex \cite{DBLP:journals/iandc/CeceF05}]
\label{def:orig-half-duplex}
A CSM is called \emph{half-duplex} if for each reachable configuration and for every pair of processes $\procA$ and $\procB$, at most one of the channels $\channel{\procA}{\procB}$ and $\channel{\procB}{\procA}$ is non-empty.
\end{definition}

Let us recall a lemma that summarises properties of execution prefixes of~CSMs.

\begin{lemma}[\cite{DBLP:conf/concur/MajumdarMSZ21}, Lemma 19]
\label{lm:execution-prefix-and-channel-content}
Let $\CSM{A}$ be a CSM.
Then, for any run $(q_0, \xi_0) \xrightarrow{w_0} \ldots \xrightarrow{w_n} (q,\xi)$ with trace $w = w_0\ldots w_n$,
it holds that (1) $\xi(\channel{\procA}{\procB}) = u$
where $\MsgVals(w \wproj_{\snd{\procA}{\procB}{\_}}) = \MsgVals(w \wproj_{\rcv{\procA}{\procB}{\_}}).u$
for every pair of processes $\procA, \procB \in \Procs$
and
(2) $w$ is \channelcompliant.
Maximal traces of $\CSM{A}$ are \channelcompliant and complete.
\label{lm:csm-channelcompliant-traces-only}
\end{lemma}

\noindent Intuitively, the lemma relates all events a CSM has executed and its current channel contents.
With this, the equivalence of both definitions follows directly.

\begin{corollary}
Both definitions of half-duplex communication, i.e., \cref{def:half-duplex} and \cref{def:orig-half-duplex}, are equivalent.
\end{corollary}

 \subsection{Existence of Linearisations Entails Causal Delivery}
\label{app:causal-delivery-ex-sync}

\textbf{Proof of \cref{lm:msc-sat-causal-delivery}. }
Let $w = e_1 \ldots$ be a sequence of events.
We claim that $w$ is the witness for causal delivery of $\msc(w)$.
Let $e_i \leq_{\msc(w)} e_j$ be two distinct events such that $e_i = \snd{\procA}{\procB}{\_}$ and $e_j = \snd{\procA}{\procB}{\_}$.
Notice that $i < j$ since $w$ is a linearisation of $\msc(w)$.
We do a case analysis whether $e_j$ is matched in $w$.
Suppose that $e_j$ is unmatched in $w$, then causal delivery holds.
Suppose that $e_j$ is matched in $w$.
Then, there is some $e_{j'}$ with $e_j \leq_{\msc(w)} e_{j'}$ and thus $j < j'$ such that $e_j \match e_{j'}$.
By definition, it holds that
\[
 \MsgVals((e_1 \ldots e_{j'}) \wproj_{\rcv{\procA}{\procB}{\_}})
 =
 \MsgVals((e_1 \ldots e_j) \wproj_{\snd{\procA}{\procB}{\_}}).
\]
We know that
$\MsgVals((e_1 \ldots e_i) \wproj_{\snd{\procA}{\procB}{\_}})$
$\preforder$
$\MsgVals((e_1 \ldots e_j) \wproj_{\snd{\procA}{\procB}{\_}})$.
Therefore, there is a prefix
$\MsgVals((e_1 \ldots e_{i'}) \wproj_{\rcv{\procA}{\procB}{\_}})$
of the sequence
$\MsgVals((e_1 \ldots e_{j'}) \wproj_{\rcv{\procA}{\procB}{\_}})$
for some $i'$
such that
$\MsgVals((e_1 \ldots e_{i'}) \wproj_{\rcv{\procA}{\procB}{\_}})$
$=$
$\MsgVals((e_1 \ldots e_i) \wproj_{\snd{\procA}{\procB}{\_}})$.
Therefore, it holds that  $e_{i'} \leq_{\msc(w)} e_{j'}$.
For causal delivery, it remains to show that $i < i'$.
Towards a contradiction, suppose that $i' \leq i$.
Since every event either represents a send or receive event, it cannot hold that $i' = i$ and therefore $i' < i$.
However, in combination with
$\MsgVals((e_1 \ldots e_{i'}) \wproj_{\rcv{\procA}{\procB}{\_}})$
$=$
$\MsgVals((e_1 \ldots e_i) \wproj_{\snd{\procA}{\procB}{\_}})$,
this renders the conditions in the definition of $\msc(\hole)$ (\cref{lm:msc-function}), for the prefix $e_1 \ldots e_{i}$ unsatisfied and thus $\msc(w)$ would be undefined which is a contradiction.
 \hfill \qed
  \section{Supplementary Material on HMSCs for \cref{sec:hmsc}}

\subsection{Complete \Channelcompliant Words}

It is easy to check that HMSCs specify complete \channelcompliant words.

\begin{proposition}
\label{prop:msc-channelcompliant}
The language $\lang(H)$, respectively $\lang(M)$, for any HMSC $H$, respectively MSC $M$, is a set of complete \channelcompliant words.
\end{proposition}

\subsection{Every HMSC is Existentially $B$-bounded for some $B$}
\label{app:proof-hmsc-exis-bounded}

\textbf{Proof of \cref{lm:HMSC-ex-B-bounded}. }
Given a single BMSC $M'$, half the number of events is a straightforward bound on channels.
One can also compute a tightest bound $B$ for which $M'$ is existentially $B$-bounded.
For more details, we refer to work by Genest et al. \cite[Section 3.3]{DBLP:journals/fuin/GenestKM07} where they define a linearisation function $\operatorname{OPT}(\hole)$ for MSCs similar to $\qOPT(\hole)$ but optimised for least channel bound.
For HMSCs, any $H = (V, \edges, v^I, V^T, \mu)$ is constructed using a finite number of BMSCs in $\mu(V)$.
Consider the maximum bound $B$ of all individual bounds for BMSCs in $\mu(V)$.
We claim that $H$ is existentially $B$-bounded
so we need to show that every MSC $M$ of $H$ is existentially $B$-bounded.
By construction, $M$ is the concatenation of (a possibly infinite number of) BMSCs $M'_1 \ldots$.
We can construct a linearisation $w = w_1 \ldots$ of $M$ such that $w_i$ is a linearisation of $M'_i$ for every $i$.
By definition of BMSCs, there is a receive event for every send event and hence all channels are empty after $w_1 \ldots w_i$ for every $i$.
Combining these observations yields that $w$ is $B$-bounded and therefore $M$ is existentially $B$-bounded.
 \hfill \qed

\subsection{Every 1-synchronisable HMSC is Half-duplex}
\label{app:proof-exis-1-sync-half-duplex}

\textbf{Proof of \cref{lm:exis-1-sync-hmsc-half-duplex}. }Let $H$ be an $1$-synchronisable HMSC.
We show that $\lang(H)$ is half-duplex.
To this end, it suffices to show that $\lang(M)$ for every MSC $M$ of $H$ is half-duplex.
Let $M = (\eventnodes, p, f, l, (\leq_\procA)_{\procA \in \Procs })$ be an MSC of $H$.
By assumption, $M$ is $1$-synchronisable.
Let $\bar{w}$ be the linearisation of $M$ from \cref{def:ex-k-synchronous}.
Since $M$ is an MSC and the way $\bar{w}$ is chosen as witness for $1$-synchronisability, every send event is immediately followed by its corresponding receive event in $\bar{w}$.

Let $w \in \lang(M)$ be any word.
We show that $w$ is half-duplex.
Towards a contradiction, suppose that the channel from $\procB$ to $\procA$ is not empty and $\procA$ attempts to send a message to $\procB$ after the prefix $w_1 \ldots w_j$ of $w$:
$w_{j+1} = \snd{\procA}{\procB}{\_}$ and 
$
\MsgVals(w_1 \ldots w_j \wproj_{\rcv{\procB}{\procA}{\_}}) 
\leq
\MsgVals(w_1 \ldots w_j \wproj_{\snd{\procB}{\procA}{\_}}) 
$.
Thus, there is $i \leq j$ such that 
$w_i = \snd{\procB}{\procA}{\_}$ which is unmatched.
By definition, the per-process order \mbox{$\leq_M \inters \; (\inv{p}(\procA) \times \inv{p}(\procA))$} is total for every process~$\procA$.
In the linearisation $\bar{w}$, the corresponding receive event happens directly after $w_i$ so the next event by~$\procA$ must be the corresponding reception which contradicts the assumption that $w_{j+1}$ is a send event of $\procA$.
 \hfill \qed
 \section{Supplementary Material on Multiparty Session Types \\ and the Indistinguishability Relation for \cref{sec:mst}}

\subsection{Semantics of Global Types from MSTs}
\label{app:semantics-mst}

\begin{definition}[\cite{DBLP:conf/concur/MajumdarMSZ21}, Def.~10: Type language for global types]
\label{def:language-global-mst} 
The type language of a global type $G$ is given as language of a finite state machine $\semglobal(G)$.
To this end, an auxiliary state machine is defined as $M(G) = (Q_{M(G)}, \Sigma_{\mathit{sync}}, δ_{M(G)}, q_{0, M(G)}, F_{M(G)})$ with:
\begin{itemize}
\item $Q_{M(G)}$ is a set consisting of all syntactic subterms in $G$ and the term $0$,
\item $δ_{M(G)}$ is the smallest set containing \\
            $(\sum_{i ∈ I} \msgFromTo{\procA}{\procB_i}{\val_i.G_i}, \msgFromTo{\procA}{\procB_i}{\val_i}, G_i)$ for each $i ∈ I$,
            as well as \\
            $(μ t. G', ε, G')$ and $(t, ε, μ t. G')$ for each subterm~$\mu t.G'$ of~$G$,
\item $q_{0, M(G)} = G$, and
\item $F_{M(G)} = \set{0}$.
\end{itemize}

Each message $\msgFromTo{\procA}{\procB}{\val}$ of $\Alphabet_{\mathit{sync}}$ is split into two events: $\snd{\procA}{\procB}{\val}$ followed by $\rcv{\procA}{\procB}{\val}$, using $\expansion(M) = (Q_{\expansion(M)}, \Alphabet, \delta_{\expansion(M)}, q_{0, \expansion(M)}, F_{\expansion(M)})$ which is defined as follows when input a state machine $M = (Q_M, \Alphabet_{\mathit{sync}}, \delta_M, q_{0 M}, F_M)$:
\begin{itemize}
\item $Q_{\expansion(M)} = Q_{M} ∪ (Q_{M} × Σ_{\mathit{sync}} × Q_{M})$,
\item $δ_{\expansion(M)}$ is the smallest set containing the transitions
\\
  $(s,\snd{\procA}{\procB}{\val},(s,\msgFromTo{\procA}{\procB}{\val},s'))$
  and $((s,\msgFromTo{\procA}{\procB}{\val},s'),\rcv{\procA}{\procB}{\val},s'))$
  \\
  for each transition $(s,\msgFromTo{\procA}{\procB}{\val},s') ∈
  δ_{M}$,
\item $q_{0, \expansion(M)} = q_{0 M}$,
and
\item $F_{\expansion(M)} = F_{M}$.
\end{itemize}
The finite state machine $\semglobal(G) \is \expansion(M(G))$ gives the type language $\lang(\semglobal(G))$ of any global type $G$.
It is straightforward that the semantics specify complete \channelcompliant sequences of events.
For brevity, $\lang(G)$ denotes $\lang(\semglobal(G))$.
\end{definition}
 \subsection{HMSCs Closed under Indistinguishability Relation}
\label{app:hmsc-indistinguishability-relation}

\textbf{Proof of \cref{lm:hmsc-closed-indistinguishability-relation}. }
We prove the claim by two inclusions.
The first inclusion $\lang(H) \subseteq \interswaplang(\lang(H))$ trivially holds.
We show that $\interswaplang(\lang(H)) \subseteq \lang(H)$.

We first prove a claim that we will use for both $\interswaplang(\lang(H)) \inters \Alphabet^*$ and
$\interswaplang(\lang(H)) \inters \Alphabet^\omega$.

\emph{Claim I:}
Let $M$ be some MSC of $H$.
Let $w$ be a sequence of events such that there is $w' \in \lang(M)$ with $w \interswap_1 w'$.
Then, $w \in \lang(M)$.

\emph{Proof of Claim I.}
We do a case analysis on the rule of $\interswap$ which has been applied to obtain \mbox{$w \interswap_1 w'$}.
In all cases, it is crucial to observe that two consecutive events are swapped and hence transitive dependencies in $\leq_M$ cannot kick in and the events either have to be ordered by the total process orders or constitute send-reception pairs.
\begin{enumerate}[(1)]
 \item Here, two send events for different processes are swapped. These two events cannot be ordered by~$\leq_M$ (without some intermediate receive event which is not present) and therefore $w \in \lang(M)$.
 \item In this case, two receive events for different processes are swapped. Again, they cannot be ordered by $\leq_M$ without some intermediate send event and the claim follows.
 \item This case deals with swapping a send event $\snd{\procA}{\procB}{\val
}$ and a receive event $\rcv{\procC}{\procD}{\val'}$.
 The first condition $\procA \neq \procD$ ensures that both events do not belong to the same process. However, this does not suffice.
 We do a case split according to the disjunction of the conditions. \\
 First, $\procA \neq \procC$ entails that the sender of the send event is neither the sender nor the receiver of the receive event.
 Then, both events cannot be related by the process order but they do also not constitute a send-reception pair (ordered through $f$) and hence the claim follows. \\
 Second, $\procB \neq \procD$ entails that the receiver of the send event and the receiver of the receive event are not the same.
 This ensures that we do not swap a reception before the corresponding send event. Again, they cannot be ordered by~$\leq_M$.
 \item This case also deals with swapping a send event $\snd{\procA}{\procB}{\val}$ and a receive event $\rcv{\procA}{\procB}{\val'}$.
 Note that $\val$ might be the same as $\val'$ and the side condition is needed to ensure that both do not consistitute a pair of send and receive event so that they were ordered through $f$.
\end{enumerate}
\emph{End Proof of Claim I.}

We first consider the case of finite sequences of events.
Let $u \in \interswaplang(\lang(H)) \inters \Alphabet^*$.
We show that $u \in \lang(H)$.
By definition, there is an MSC $M$ of $H$ such that $u \in \interswaplang(\lang(M)) \inters \Alphabet^*$.
Therefore, there is $u' \in \lang(M)$ such that $u' \interswap_n u$ for some~$n$.
By applying \emph{Claim I} $n$-times, the claim follows.

Second, let $u \in \interswaplang(\lang(H)) \inters \Alphabet^\omega$.
By definition, there is an MSC $M$ of $H$ such that $u \in \interswaplang(\lang(M)) \inters \Alphabet^\omega$.
Therefore, there is $v \in \lang(M)$ such that $u \preceq^\omega_\interswap v$ which means that for every prefix $u'$ of~$u$, there is some prefix $v'$ of~$v$ such that $u' \preceq v'$, i.e., there is some $w$ such that $u'w \interswap v'$.
From the case for finite sequences, it follows that $u' \in \pref(\lang(M))$ for every finite prefix $u'$ of~$u$.
Note that there is a single (infinite) MSC M for which this applies.
Therefore, the path of every prefix $u'$ in $H$ is the same and can be extended for longer prefixes.
Since every prefix $u'$ of~$u$ is in $\pref(\lang(M))$, the infinite sequence $u$ is in $\lang(M)$ and hence in~$\lang(H)$.
\hfill \qed
  \subsection{Indistinguishability Relation Preserves \\ Satisfaction of Channel Restrictions for Channel-compliant Words}
\label{app:protocols-closed-indistinguishability-relation}

\textbf{Proof of \cref{lm:protocols-closed-indistinguishability-relation}. }
Let $w$ be a complete \channelcompliant word.

First, suppose that $w$ is half-duplex.
It is straightforward to check that $\interswap$ does not swap any two events $\rcv{\procA}{\procB}{\val}$ and $\snd{\procB}{\procA}{\val'}$ and hence the condition for half-duplex is preserved.

Second, suppose that $w$ is existentially $B$-bounded for some $B$.
Then, we know that $\msc(w)$ is a prefix MSC which admits a $B$-bounded linearisation.
Analogous to the proof of \cref{lm:hmsc-closed-indistinguishability-relation}, we can show that any (even infinite) prefix MSC is closed under $\interswap$.
Therefore, for any $w'$ for which $w' \interswap w$, it holds that $\msc(w) = \msc(w')$ and the latter still admits the same $B$-bounded linearisation.

Third, suppose that $w$ is $k$-synchronisable.
Recall that $k$-synchronisability is also defined on $\msc(w)$ and hence the same reasoning as for the second case applies.
\hfill \qed
 \subsection{Correctness of the Embedding of Global Types from MSTs into HMSCs}
\label{app:relation-mst-hmsc}

Prior to giving the correctness statement of the embedding, we define a translation from HMSCs of special shape to FSMs.

\begin{definition}
 We say that an HMSC $H = (V, \edges, v^I\negmedspace, V^T\negmedspace\!, \mu)$ is a \emph{$1$-HMSC} iff every BMSC in $\mu(V)$ consists of at most one send and one receive event.
\end{definition}

\begin{definition}[Quasi-optimal translation of $H$]
\label{def:qOPT-for-HMSC}
Let $H = (V,\edges,v^I\negmedspace,V^T\negmedspace\!,μ)$ be a HMSC such that
for every $v \in V$, $\mu(v) = M_\emptyset$ or $\mu(v) = M( \msgFromTo{\procA}{\procB}{\val} )$ for some $\procA$, $\procB$ and $\val$.
We define the function $\qOPT(M)$ if $M$ is either
$M_\emptyset$ or
$M( \msgFromTo{\procA}{\procB}{\val} ):$
$\qOPT(M_\emptyset) = \emptystring$ and
$\qOPT(M( \msgFromTo{\procA}{\procB}{\val} )) = \msgFromTo{\procA}{\procB}{\val}$.
The state machine $\qOPT(H(G)) = (Q,Σ_G,δ,q_0,F)$ is defined as follows.

\medskip
{\centering
$
 Q  =  \set{\vertexA¹, \vertexA² \mid v∈V} \hfill
q₀ =  \set{v¹ \mid v=v^I} \hfill
 F  =  \set{v² \mid v∈V^T}
$
}
\[
 δ  =  \set{ (v¹, \qOPT(μ(v)), v²) \mid v∈V ∧  \qOPT(μ(v)) ∈  Σ_{\mathit{sync}} \union \set{ε} }
        \; ∪ \;  \set{ (v²,u¹) \mid (v,u) ∈ E }
\]
\end{definition}

Note that the above conditions are satisfied by $H(G)$ for any global type $G$.
Recall that we defined $\expansion(\hole)$ in \cref{def:language-global-mst} to expand the alphabet $\Alphabet_{\mathit{sync}}$ to~$\Alphabet$, i.e., to split into send and receive events.

\begin{lemma}[Correctness of \cref{def:qOPT-for-HMSC}]
\label{lm:correctness-qOPT}
Let $H$ be a $1$-HMSC.
Then,
$\lang(\expansion(\qOPT(H))) \subseteq \lang(H)$ and
$\interswaplang(\lang(\expansion(\qOPT(H)))) = \lang(H)$.
\end{lemma}
\textbf{Proof. }The expansion operator $\expansion(\hole)$ is a technical mean to let two alphabets match.
However, in case we do not apply the indistinguishability relation, it does not matter and we will waive $\expansion(\hole)$ in our proofs for readability.

For the first claim, we prove the following:

\emph{Claim I:}
Let $w$ be some word with the following run in $\qOPT(H)$: \\ 
\phantom{some text to move to the right} $\vertexA_1^1, \vertexA_1^2, \vertexA_2^1, \vertexA_2^2, \ldots, \vertexA_n^1, \vertexA_n^2$. \\
We claim that
$w \in \lang(\mu(\vertexA_1)\mu(\vertexA_2) \cdots \mu(\vertexA_n))$
with the same vertices.

\emph{Proof of Claim I:}
We prove this by induction on the number of vertices~$n$.
Note that, as $\mu(v_i)$ can be $M_\emptyset$ for some $i$, this is not necessarily an induction on $\card{w}$.
For the base case, let $n = 0$.
Then, the word is empty and the claim trivially holds.
For the induction step, let us assume that the claim holds for $n$ and~$w$.
We prove it for $n+1$ and $w'$.
By construction $w' = wx$ where $x$ is either $\emptystring$ or $\msgFromTo{\procA}{\procB}{\val}$ (respectively $\snd{\procA}{\procB}{\val}.\rcv{\procA}{\procB}{\val}$ in the expansion).
The run in $\qOPT(H)$ is
$\vertexA_1^1, \vertexA_1^2, \ldots, \vertexA_n^1, \vertexA_n^2, \vertexA_{n+1}^1, \vertexA_{n+1}^2$.
By construction, $x \in \lang(\mu(v_{n+1}))$.
Hence, $wx \in \lang(\mu(v_1) \cdots \mu(v_{n+1}))$ which proves Claim I.

\emph{End Proof of Claim I.}\\
Claim I shows the inclusion for the finite case when $\vertexA_n$ is final:
\[
\lang(\expansion(\qOPT(H))) \inters \Alphabet^*
\subseteq
\lang(H) \inters \Alphabet^*.
\]
It remains to show the infinite case:
$\lang(\expansion(\qOPT(H))) \inters \Alphabet^\omega
\subseteq
\lang(H) \inters \Alphabet^\omega$.

Let
$w \in \lang(\expansion(\qOPT(H))) \inters \Alphabet^\omega$ be some infinite word.
We show that there is some infinite path $\vertexA_1, \vertexA_2, \ldots$ in $H$ with linearisation $w$.
Consider a tree $\mathcal{T}$ where each node corresponds to some path $\pi$ in~$H$ whose linearisations are prefixes $w'$ of $w$.
The root is labelled by the empty path.
The children of a node $\pi$ are paths that extend $\pi$ by a single node ---
these exist from the reasoning in the induction step for \emph{Claim I}.
HMSC $H$ is finitely branching and so is $\mathcal{T}$.
By König's Lemma, there is an infinite path in $\mathcal{T}$ whose linearisation is $w$ which concludes the proof of the first claim.

We prove the second claim by proving two inclusions.
The first inclusion
$\interswaplang(\lang(\expansion(\qOPT(H)))) \subseteq \lang(H)$ follows from the first claim and \cref{lm:hmsc-closed-indistinguishability-relation}.
For the second inclusion, we show the following:

\emph{Claim II:}
Let $\pi = \vertexA_1, \ldots, \vertexA_n$ be some path in $H$.
For all sequences $w \in \pref(\lang(\mu(\vertexA_1) \cdots \mu(\vertexA_n)))$, there is some $w' \in \lang(\expansion(\qOPT(H)))$ with some run $\vertexA_1^1, \vertexA_1^2, \ldots, \vertexA_n^1, \vertexA_n^2$ such that $w \preceq w'$.

\emph{Proof of Claim II:}
We prove this by induction on the number of vertices $n$.
For the base case, let $n = 0$.
Then, the word is actually empty and the claim trivially holds.
For the induction step, let us assume that the claim holds for $n$ with $w_n$ and $w'_n$ and we prove it for $n+1$ with $w_{n+1}$ and $w'_{n+1}$.
We know that $w_{n+1} \in \lang(\mu(\vertexA_1) \cdots \mu(\vertexA_{n+1}))$.
By definition of the language operator for HMSCs, $w_{n+1} \interswap w_n.x \; (*)$ such that $w_n \in \lang(\mu(\vertexA_1) \cdots \mu(\vertexA_n))$ and $x \in \lang(\mu(\vertexA_{n+1}))$.
For $w_n$, there is $w'_n$ with the run 
$\vertexA_1^1, \vertexA_1^2, \ldots, \vertexA_n^1, \vertexA_n^2$ 
in $\expansion(\qOPT(H))$.
By construction, we can extend this run for $w'_n.x$ as follows:
$\vertexA_1^1, \vertexA_1^2, \ldots, \vertexA_n^1, \vertexA_n^2, \vertexA_{n+1}^1, \vertexA_{n+1}^2$.
By the induction hypothesis and~$(*)$,
$w_{n+1} \preceq w'_n.x$,
which concludes the proof.

\emph{End Proof of Claim II.}

We prove that $\lang(H) \subseteq \interswaplang(\lang(\expansion(\qOPT(H))))$.
Let $w \in \lang(H)$.

In case $w$ is finite, there is a finite path $\pi = \vertexA_1, \ldots, \vertexA_n$ for which $w \in \lang(\mu(\vertexA_1), \ldots, \mu(\vertexA_n))$ where $\vertexA_n$ is final.
We apply Claim II and know that $\vertexA_n^2$ is also final.
We also know that $w \preceq w'$ for some $w'$ with a run through the corresponding states in the automaton.
By construction, $w'$ cannot be longer than $w$ and therefore $w \interswap w'$ which proves the claim.

In case $w$ is infinite, there is a infinite path $\pi = \vertexA_1, \ldots$ for which $w \in \lang(\mu(\vertexA_1, \ldots))$.
For every prefix $u$ of~$w$, there is $n$ such that $u \in \pref(\lang(\mu(\vertexA_1) \cdots \mu(\vertexA_n)))$.
By Claim II, there is $u' \in \lang(\expansion(\qOPT(H)))$ with run $\vertexA_1^1, \vertexA_1^2, \ldots, \vertexA_n^1, \vertexA_n^2$ such that $u \preceq u'$.
Consider a tree $\mathcal{T}$ where each node corresponds to a prefix of the run $\vertexA_1^1, \vertexA_1^2, \ldots$ in $\expansion(\qOPT(H))$.
The root is labelled by the empty run.
The children of a node $\rho$ are the runs that extends $\rho$ by a single transition -- which exist by the above reasoning.
Since $\mathcal{T}$ is finitely branching, there is an infinite path in $\mathcal{T}$ that corresponds to an infinite run $\rho$ in $\expansion(\qOPT(H))$ for which $w \preceq^\omega \trace(\rho)$ which concludes the proof.
\hfill \qed

With weak bisimulations, we show that $\lang(M(G))$ and $\lang(\qOPT(H(G)))$ yield the same language for any global type~$G$.

\begin{definition}[Weak bisimulation]
Let $A = (Q_A, \Alphabet, \delta_A, q_{0, A}, F_A), B = (Q_B, \Alphabet, \delta_B, q_{0, B}, F_B)$ be two state machines.
We say that $R \subseteq Q_A \times Q_B$ is a \emph{weak simulation relation} for $A$ and $B$ is relation with the following properties:
\begin{itemize}
 \item for all $a \in \Alphabet, s_A, t_A \in S_A, s_B \in S_B$,
        if $R(s_A, s_B)$ and $(s_A, a, t_A) \in \delta_A$,
        then there is a $t_B \in Q_B$ such that
        $(s_B, a, t_B) \in \delta_B^*$,
 \item for all $s_A, t_A \in S_A, s_B \in S_B$,
        if $R(s_A, s_B)$ and $(s_A, \emptystring, t_A) \in \delta_A$,
        then there is a $t_B \in Q_B$ such that
        $(s_B, \emptystring, t_B) \in \delta_B^*$,
 \item $R_1(q_{0, A}, q_{0, B})$ and $R_2(q_{0, B}, q_{0, A})$, and
 \item for every $q_A \in F_A$, there is a $q_B \in F_B$ such that $R_1(q_A, q_B)$ and vice versa.
\end{itemize}
\end{definition}

This definition is given as sufficient condition for the original definition of weak simulation by Milner \cite[Def.~6.2 and Prop.~6.3]{DBLP:books/daglib/0098267}.
For clarity in our context, we amended it to refer to two different sets of states instead of merging both to a single set of states.

\begin{proposition}
\label{lm:weakbisimIMPlangequiv}
Let $A = (Q_A, \Delta, \delta_A, q_{0, A}, F_A), B = (Q_B, \Delta, \delta_B, q_{0, B}, F_B)$ be two state machines.
Let $R_1 \subseteq Q_A \times Q_B$ and $R_2 \subseteq Q_B \times Q_A$ be two weak bisimulations for $A$ and $B$ (respectively $B$ and $A$).
Then, $\lang(A) = \lang(B)$.
\end{proposition}

\begin{lemma}
\label{lm:weakbisimExists}
Let $G$ be a global type, 
$M(G)$ the corresponding state machine with state space $Q = \set{q_{G'} \mid G' \text{ is syntactic subterm of } G}$ 
and 
$\qOPT(\mpstTOhmscc(G))$ the state machine built from the HMSC $\mpstTOhmscc(G)$ with states $V = \set{v_{G'}^i \mid i \in \set{1,2} \text{ and } G' \text{ is a syntactic subterm of } G}$ as defined before.
Recall that there is one state and two vertices for every syntactic subterm of $G$.
Hence, we index states $q$ and vertices $\vertexA$ by these subterms.
For the auxiliary vertices in $\qOPT(\mpstTOhmscc(G))$, we use $G^j\!$.
There are two weak bisimulation relations 
$R₁ ⊆  Q \times V$ and $R₂ ⊆  V \times Q$ s.t. 
\begin{enumerate}[(1)]
 \item $R₁(q_G, \vertexA¹_G)$ and $R₂(\vertexA¹_G, q_G)$ as well as
 \item $R₁(q_0, \vertexA²_0)$ and $R₂(\vertexA²_0, q_0)$.
\end{enumerate}
\end{lemma}
\textbf{Proof. }
We prove the claim by structural induction on $G$.
\begin{itemize}
 \item Base $G = 0$ : \\
       There is one state in $M(G)$ and two vertices in $\qOPT(\mpstTOhmscc(G))$.
         It is straightforward to define $R₁$~and $R₂$ such that the conditions are satisfied.
         Both relations are weak bisimulation relations as there is a solely one $ε$-transition.
 \item Step $G = \Sum_{i ∈ I} \msgFromTo{\procA}{\procB_i}{\val_i}.G_i$: \\
         The induction hypothesis holds for every $i ∈  I$ with $R₁^i$ and $R₂^i$.
         We define:
         \begin{align*}
            R₁ \is & ⋃_{i∈ I} R₁^i ∪ \set{(q_G, \vertexA¹_G), (q_G, \vertexA²_G)} \text{, and} \\
            R₂ \is & ⋃_{i∈ I} R₂^i ∪ \set{(\vertexA¹_G, q_G), (\vertexA²_G, q_G)} \; \union \;
            ⋃_{i∈ I} \set{(\vertexA¹_{G^i}, q_{G_i}), (\vertexA²_{G^i}, q_{G_i})}
         \end{align*}
         By this, (1) and (2) are satisfied.
         For the remaining conditions of a weak bisimulation relation, it suffices to check the added states (and states with new transitions) since the induction hypotheses are sufficient for the rest. \\
\textit{Added states}: $q_G$ as well as $\vertexA¹_G, \vertexA²_G, \vertexA¹_{G^i}, \vertexA²_{G^i}$ for every $i ∈  I$. \\
         \textit{Added transitions}:
         $(q_G, \msgFromTo{\procA}{\procB_i}{\val_i}, q_{G_i})$
                                for every $i ∈  I$
                          and $(\vertexA¹_G, ε, \vertexA²_G)$,
                              $(\vertexA²_G, ε, \vertexA¹_{G^i})$, 
                              $(\vertexA¹_{G^i}, 
                              \msgFromTo{\procA}{\procB_i}{\val_i}, \vertexA²_{G^i})$, and
                              $(\vertexA²_{G^i}, ε, \vertexA¹_{G_i})$.  \\
         We check $R₁$ and hence $q_G$:
         \begin{itemize}
          \item $l ∈  Σ_{\mathit{sync}}$ is the only possible transition 
                    from $q_G$ to $q_{G_i}$ for every $i ∈  I$; 
            it suffices to check $\vertexA¹_G$ as the other state $\vertexA²_G$ that is in relation with $R₁$ occurs on the path.
            $R₁(q_G, \vertexA¹_G)$ and $\vertexA¹_G \overset{ε}{→} \vertexA²_G \overset{ε}{→} \vertexA¹_{G^i} \overset{l}{→} \vertexA²_{G^i} \overset{ε}{→} \vertexA¹_{G_i}$.
            This is fine by the induction hypotheses: $R₁^i(q_{G_i}, \vertexA¹_{G_i})$.
            There are no $ε$-transitions from $q_G$.
         \end{itemize}
         We check $R₂$ and hence $\vertexA¹_G, \vertexA²_G, \vertexA¹_{G^i}, \vertexA²_{G^i}$ for every $i ∈  I$:
         \begin{itemize}
          \item $\vertexA¹_G$: \\
                One can only take an $ε$-transition from $\vertexA¹_G$, i.e.,
                $\vertexA¹_G \overset{ε}{→}  \vertexA²_G$; 
                by definition $R₂(\vertexA²_G, q_G); R₂(\vertexA¹_G, q_G)$ and $*$ allows not to move.
          \item $\vertexA²_G$: \\
                One can only take an $ε$-transition from $\vertexA²_G$, i.e.,
                $\vertexA²_G \overset{ε}{→}  \vertexA¹_{G^i}$ for every $i ∈  I$;
                by definition $R₂(v_{G^i}, q_G); R₂(v_G, q_G)$ and $*$ allows not to move.
          \item $\vertexA¹_{G^i}$: \\
                One can only take a non-$ε$-transition from $\vertexA¹_{G^i}$, i.e.,
                $\vertexA¹_{G^i} \overset{l}{→} \vertexA²_{G^i}$; 
                by definition $R₂(\vertexA¹_{G^i}, q_G),$ $R₂(\vertexA²_{G^i}, q_{G_i})$ and $q_G \overset{l}{→} q_{G_i}$ and hence fine.
          \item $\vertexA²_{G^i}$: \\
                One can only take an $ε$-transition from $\vertexA²_{G^i}$ to $\vertexA¹_{G_i}$; 
                and by definition $R₂(\vertexA²_{G^i}, q_{G_i}),$
                $R₂(\vertexA¹_{G_i}, q_{G_i})$ by the induction hypotheses 
                and $*$ allows not to move.
         \end{itemize}
 \item Step $G = μ t. G'$: \\
         The induction hypothesis holds for $G'$ with $R₁'$ and $R₂'$.
         Recall that $Q = \set{ q_{G'} \mid G' ⊑ G}$.
         We define:
         \begin{align*}
            R₁ \is \; & R₁' ∪ \set{(q_G, \vertexA¹_G), (q_G, \vertexA²_G)} \text{, and} \\
            R₂ \is \; & R₂' ∪ \set{(\vertexA¹_G, q_G), (\vertexA²_G, q_G)}.
         \end{align*}
         By this, (1) and (2) are already satisfied. 
         For the conditions for a weak bisimulation relation, it suffices to check the added states and states with new transitions. \\
         \textit{Added states}:        $q_G$ 
                   as well as $\vertexA¹_G$ and $\vertexA²_G$. \\
         \textit{Added transitions}:
         $ 
(\vertexA¹_G, ε, \vertexA²_{G'}), (\vertexA²_G, ε, \vertexA¹_{G'}), (\vertexA²_t, ε, \vertexA¹_G)
         $ 
         as well as
         $ 
            (q_G, ε, q_{G'}), (q_t, ε, q_G).
         $ 
         
         The induction hypotheses apply for all other states and transitions.
         We check $R₁$ and hence $q_G$:
         \begin{itemize}
          \item One cannot take a non-$ε$-transition from $q_G$.
                The only $ε$-transition that can be taken is to $q_{G'}$;
                we only need to check $\vertexA¹_G$ as $\vertexA²_G$ is on its way and also related;
                from $\vertexA¹_G \overset{ε}{→} \vertexA²_G \overset{ε}{→} \vertexA¹_{G'}$ with an $ε$ and by induction hypothesis they are related.
         \end{itemize}
         We check $R₂$ and hence $\vertexA¹_G, \vertexA²_G$, and $\vertexA²_t$.
         \begin{itemize}
          \item $\vertexA¹_G$: \\
                Only one $ε$-transition is possible to $\vertexA^2_G$ and both are related with the same state.
          \item $\vertexA²_G$: \\
                Only one $ε$-transition is possible:
                $\vertexA²_G \overset{ε}{→} \vertexA¹_{G'}$ and $q_{G} \overset{ε}{→}  q_{G'}$
                and by definition the last two are related.
          \item $\vertexA²_t$: \\
                Only one $ε$-transition is possible:
                $\vertexA²_t \overset{ε}{→} \vertexA¹_G$
                and 
                $R₂(\vertexA²_t, q_{μt.G'})$ where $q_{μt.G'}=q_G$ and $\vertexA¹_G$ is in relation with this one 
                and taking no transition suffices.
         \end{itemize}
\end{itemize}
\hfill \qed

\begin{lemma}
\label{lm:relation-M(G)-qOPT(H(G))}
Let $G$ be a global type.
Then, $\lang(G) = \lang(\expansion(\qOPT(H(G))))$.
\end{lemma}
\textbf{Proof. }
 Recall that $\lang(G) = \lang(\semglobal(G))$ and $\semglobal(G) = \expansion(M(G))$.
 Therefore, it suffices to prove that
 $\lang(M(G)) = \lang(\qOPT(H(G)))$
 after substituting and omitting $\expansion(\hole)$.
 This follows from \cref{lm:weakbisimExists,lm:weakbisimIMPlangequiv}:
 in the former, we show that there are two weak bisimulations between both automata;
 in the latter, we state the well-known fact that, then, both languages are equivalent.
\hfill \qed

\smallskip
Equipped with this, we can now prove the correctness of the embedding of global types from MSTs into HMSCs.

\medskip
\noindent
\textbf{Proof of \cref{thm:correctnessMPSTtoHMSC}.}
Recall that $\lang(G)$ is an abbreviation for $\lang(\semglobal(G))$.
With the first fact of \cref{lm:correctness-qOPT}, it suffices to show
$\lang(\semglobal(G)) \subseteq \lang(\expansion(\qOPT(H(G))))$ for the first claim which follows with
\cref{lm:relation-M(G)-qOPT(H(G))}.

For the second claim, it suffices to show that
\[
\interswaplang(\lang(G))
\overset{\text{(A)}}{=}
\interswaplang(\lang(\expansion(\qOPT(H(G)))))
\overset{\text{(B)}}{=}
\interswaplang(\lang(H(G))).
\]
(A) follows from
\cref{lm:relation-M(G)-qOPT(H(G))} while
(B) follows from the second fact of
\cref{lm:correctness-qOPT}.
\hfill \qed
 \subsection{Languages of Global Types from MSTs are \\
half-duplex, existentially 1-bounded, and 1-synchronisable}
\label{app:mst-channel-use}

\textbf{Proof of \cref{cor:MST-ex-1-sync-and-ex-1-bounded}. }
Let $G$ be some global type.
With the correctness of our encoding (\cref{thm:correctnessMPSTtoHMSC}), it suffices to show that $\mpstTOhmscc(G)$ has the desired properties.

First, we show that $\mpstTOhmscc(G)$ is existentially $1$-bounded.
From \cref{lm:HMSC-ex-B-bounded}, we know that every HMSC~$H$ is existentially $B$-bounded for some $B$.
From the proof of \cref{lm:HMSC-ex-B-bounded}, it follows that this is the case for the maximal $B$ for which some individual BMSC of $H$ is existentially $B$-bounded.
For $\mpstTOhmscc(G)$, this maximal~$B$ is~$1$ so the claim follows.

Second, we show that $\mpstTOhmscc(G)$ is $1$-synchronisable.
We apply \cref{cor:k-sync-hmsc} and have to show that every MSC $M$ of $\mpstTOhmscc(G)$ is $1$-synchronisable.
By construction, any $M$ is the concatenation of (possibly infinitely many) BMSCs $M'_1, \ldots$ with at most one message exchange.
Therefore, it is straightforward to construct a linearisation of $M$ by concatenating the message exchanges consisting of one send and receive event of every individual BMSC.
Therefore, any MSC $M$ of $\mpstTOhmscc(G)$ is $1$-synchronisable.
This concludes the proof of the first claim.

Third, we know from \cref{lm:exis-1-sync-hmsc-half-duplex} that every $1$-synchronisable HMSC is half-duplex.
Thus, the last claim follows.
\hfill \qed

\end{document}